\documentclass[twocolumn,twocolappendix]{aastex7}
\usepackage{xspace}
\usepackage{upquote}
\usepackage{amsmath}
\usepackage{subcaption}
\usepackage{xcolor}
\usepackage{soul}
\usepackage{CJK}
\newcommand{\gaia}{\text{Gaia}\xspace}

\newcommand\gdr[1]{\gaia DR#1}

\newcommand\gbp{\ensuremath{G_\mathrm{BP}}}
\newcommand\grp{\ensuremath{G_\mathrm{RP}}}

\newcommand{\w}[1]{\text{W#1}}
\newcommand{\ks}{\ensuremath{K_\mathrm{s}}\xspace}

\graphicspath{{./}{figures/}}
\begin{document}
\begin{CJK*}{UTF8}{gbsn}
\title{The CatSouth Quasar Candidate Catalog for the Southern Sky and a Unified All-Sky Catalog Based on Gaia DR3}

\correspondingauthor{Yuming Fu}

\author[0000-0002-0759-0504]{Yuming Fu (傅煜铭)}
\email[show]{yfu@strw.leidenuniv.nl}
\affil{Leiden Observatory, Leiden University, Einsteinweg 55, 2333 CC Leiden, The Netherlands}
\affil{Kapteyn Astronomical Institute, University of Groningen, P.O. Box 800, 9700 AV Groningen, The Netherlands}

\author[0000-0002-7350-6913]{Xue-Bing Wu}
\email[show]{wuxb@pku.edu.cn}
\affil{Department of Astronomy, School of Physics, Peking University, Beijing 100871, People's Republic of China}
\affil{Kavli Institute for Astronomy and Astrophysics, Peking University, Beijing 100871, People's Republic of China}
\affil{National Astronomical Observatories, Chinese Academy of Sciences, Beijing 100101, People's Republic of China}

\author[0000-0002-4989-2471]{R. J. Bouwens}
\email{bouwens@strw.leidenuniv.nl}
\affil{Leiden Observatory, Leiden University, Einsteinweg 55, 2333 CC Leiden, The Netherlands}

\author[0000-0001-8183-1460]{Karina I. Caputi}
\email{karina@astro.rug.nl}
\affiliation{Kapteyn Astronomical Institute, University of Groningen, P.O. Box 800, 9700 AV Groningen, The Netherlands}
\affiliation{Cosmic Dawn Center (DAWN), Copenhagen, Denmark}

\author[0009-0005-3823-9302]{Yuxuan Pang}
\email{1901110223@pku.edu.cn}
\affil{Department of Astronomy, School of Physics, Peking University, Beijing 100871, People's Republic of China}
\affil{Kavli Institute for Astronomy and Astrophysics, Peking University, Beijing 100871, People's Republic of China}

\author[0000-0002-0792-2353]{Rui Zhu}
\email{rui@stu.pku.edu.cn}
\affil{Department of Astronomy, School of Physics, Peking University, Beijing 100871, People's Republic of China}
\affil{Kavli Institute for Astronomy and Astrophysics, Peking University, Beijing 100871, People's Republic of China}

\author[0000-0002-6769-0910]{Da-Ming Yang}
\email{dyang@strw.leidenuniv.nl}
\affil{Leiden Observatory, Leiden University, Einsteinweg 55, 2333 CC Leiden, The Netherlands}

\author[0000-0002-7533-1264]{Jin Qin}
\email{qinjin@stu.pku.edu.cn}
\affil{Department of Astronomy, School of Physics, Peking University, Beijing 100871, People's Republic of China}
\affil{Kavli Institute for Astronomy and Astrophysics, Peking University, Beijing 100871, People's Republic of China}

\author[0000-0001-8803-0738]{Huimei Wang}
\email{wang_hm@pku.edu.cn}
\affil{Department of Astronomy, School of Physics, Peking University, Beijing 100871, People's Republic of China}
\affil{Kavli Institute for Astronomy and Astrophysics, Peking University, Beijing 100871, People's Republic of China}

\author[0000-0002-4569-016X]{Christian Wolf}
\email{christian.wolf@anu.edu.au}
\affil{Research School of Astronomy and Astrophysics, Australian National University, Canberra ACT 2611, Australia}
\affil{Centre for Gravitational Astrophysics, Australian National University, Canberra ACT 2600, Australia}

\author[0009-0002-9096-2299]{Yifan Li}
\email{yli@strw.leidenuniv.nl}
\affil{Leiden Observatory, Leiden University, Einsteinweg 55, 2333 CC Leiden, The Netherlands}

\author[0000-0002-5535-4186]{Ravi Joshi}
\email{rvjoshirv@gmail.com}
\affil{Indian Institute of Astrophysics, Koramangala, Bangalore 560034, India}

\author[0000-0002-6610-5265]{Yanxia Zhang}
\email{zyx@bao.ac.cn}
\affil{National Astronomical Observatories, Chinese Academy of Sciences, Beijing 100101, People's Republic of China}

\author[0009-0003-3066-2830]{Zhi-Ying Huo}
\email{zhiyinghuo@bao.ac.cn}
\affil{National Astronomical Observatories, Chinese Academy of Sciences, Beijing 100101, People's Republic of China}

\author{Y. L. Ai}
\email{aiyanli@sztu.edu.cn}
\affil{College of Engineering Physics, Shenzhen Technology University, Shenzhen 518118, People's Republic of China}
\affil{Shenzhen Key Laboratory of Ultraintense Laser and Advanced Material Technology, Shenzhen 518118, People's Republic of China}

\begin{abstract}

The Gaia DR3 has provided a large sample of more than 6.6 million quasar candidates with high completeness but low purity. Previous work on the CatNorth quasar candidate catalog has shown that including external multiband data and applying machine-learning methods can efficiently purify the original Gaia DR3 quasar candidate catalog and improve the redshift estimates. In this paper, we extend the Gaia DR3 quasar candidate selection to the southern hemisphere using data from SkyMappper, CatWISE, and VISTA surveys. We train an XGBoost classifier on a unified set of high-confidence stars and spectroscopically confirmed quasars and galaxies. For sources with available Gaia BP/RP spectra, spectroscopic redshifts are derived using a pre-trained convolutional neural network (RegNet). We also train an ensemble photometric redshift estimation model based on XGBoost, TabNet, and FT-Transformer, achieving an RMSE of 0.2256 and a normalized median absolute deviation of 0.0187 on the validation set. By merging CatSouth with the previously published CatNorth catalog, we construct the unified all-sky CatGlobe catalog with nearly 1.9 million sources at $G<21$, providing a comprehensive and high-purity quasar candidate sample for future spectroscopic and cosmological investigations.

\end{abstract}



\section{Introduction} \label{sec:intro}

Quasars are luminous active galactic nuclei (AGNs) powered by accretion onto supermassive black holes. Observable across vast cosmic distances, quasars are ideal probes for many astrophysical and cosmological studies. Comprehensive quasar samples enable investigations into the formation and evolution of supermassive black holes \citep[e.g.][]{2015Natur.518..512W,2018Natur.553..473B,2020ARA&A..58...27I,2023ARA&A..61..373F}, the co-evolution of black holes and galaxies \citep[e.g.][]{2005Natur.433..604D,2013ARA&A..51..511K}, and the structure and composition of the intergalactic medium \citep[e.g.][]{1981ARA&A..19...41W,1986MNRAS.218P..25R,2006ApJS..165....1T}. Additionally, quasar distribution traces the large-scale structure of the Universe, providing critical constraints on cosmological models \citep[e.g.][]{2011AJ....142...72E,2013AJ....145...10D,2017AJ....154...28B}. Quasars also serve as reference sources for celestial frames with their small parallaxes and proper motions \citep[e.g.][]{2009ITN....35....1M,2016A&A...595A...5M,2018A&A...616A..14G,2022A&A...667A.148G}.

Until 2023, nearly 1 million quasars have been spectroscopically identified \citep[see e.g. The Million Quasar Catalog;][]{2023OJAp....6E..49F}. Most of these quasars are identified by the Sloan Digital Sky Surveys \citep[e.g.][]{2020ApJS..249....3A,2020ApJS..250....8L,2023ApJS..267...44A}. Other representative quasar surveys include the 2dF QSO Redshift Survey \citep[2QZ;][]{2004MNRAS.349.1397C}, the 2dF-SDSS LRG and QSO survey \citep[2SLAQ;][]{2009MNRAS.399.1755C}, the LAMOST quasar survey \citep[][]{2016AJ....151...24A,2018AJ....155..189D,2019ApJS..240....6Y,2023ApJS..265...25J}, and the Early Data Release of the Dark Energy Spectroscopic Instrument \citep[DESI EDR;][]{2024AJ....168...58D}. {{More recently, DESI Data Release 1 has published approximately 1.6 million spectroscopically classified quasars \citep{2025arXiv250314745D}.}} Nevertheless, the completeness and identification efficiency of the existing quasar samples are still restricted by factors such as the bias of candidate selection methods, instrument performance, and ground-based observation conditions.

% Unlike long-slit and multi-fiber spectroscopic quasar surveys which require candidate selections from imaging data, slitless spectroscopic surveys \citep[e.g.][]{2008A&A...487..539W} and untargeted integral field unit (IFU) surveys \citep[e.g. The Hobby-Eberly Telescope Dark Energy Experiment Survey, HETDEX;][]{2022ApJS..261...24L,2024arXiv241219414L} provide unique ways to build more complete quasar samples because the latter two do not rely on the imaging pre-selections. 

Combining (candidate) quasars selected with different methods can effectively increase sample completeness. A good example is the \gdr{3} quasar candidate catalog\footnote{The original \gdr{3} quasar candidate catalog is available at the \gaia archive \url{https://gea.esac.esa.int/archive} with table name \texttt{gaiadr3.qso\_candidates}.} \citep[hereafter GDR3 QSO candidate catalog;][]{2023A&A...674A...1G,2023A&A...674A..41G} with 6.6 million quasar candidates selected by at least one of the several different classification modules, including the Discrete Source Classifier (DSC), the Quasar Classifier (QSOC), the variability classification module, the surface brightness profile module, and the \gdr{3} Celestial Reference Frame source table. In particular, the DSC uses the \gaia BP/RP spectrum together with the mean $G$-band magnitude, the $G$-band variability, the parallax, and the proper motion to classify each \gaia source probabilistically, which is less biased than color cuts in selecting quasars. Although highly complete, the GDR3 QSO candidate catalog has low purity \citep[$\sim52\%$ as estimated by][]{2023A&A...674A..41G}, which limits its further application in astrophysical and cosmological studies.

To extract purer subsamples from the GDR3 QSO candidate catalog, approaches incorporating photometric data from external mid-infrared and optical surveys have been proposed. For example, by applying cuts on \gaia and unWISE \citep{2014AJ....147..108L} colors and \gaia proper motions to remove non-quasar contaminants (stars and galaxies), \citet{2024ApJ...964...69S} constructed the Quaia quasar catalog with nearly 1.3 million sources at $G<20.5$. Using data from \gaia, Pan-STARRS1 \citep[PS1;][]{Chambers2016}, and CatWISE2020 \citep{2021ApJS..253....8M} and machine learning methods, \citet{2024ApJS..271...54F} presented CatNorth, another improved \gdr{3} quasar candidate catalog with more than 1.5 million sources down to the \gaia limiting magnitude in the 3$\pi$ sky of PS1 footprint ($\delta>-30\degr$). \citet{2024ApJS..271...54F} have shown that the machine-learning-based CatNorth catalog has higher completeness than Quaia while obtaining similar purity. This proves that the machine learning method can better disentangle different celestial objects in the high-dimensional space than cuts on two-dimensional planes. In addition, the inclusion of PS1 photometry in \citet{2024ApJS..271...54F} also leads to higher photometric redshift accuracy in CatNorth than in Quaia. 

In contrast to the northern hemisphere, where systematic surveys such as SDSS and DESI have yielded extensive quasar catalogs covering large sky areas, the southern sky has long suffered from limited multiband coverage and less homogeneous spectroscopic follow-up. Early southern efforts, including 2QZ, the Hamburg/ESO Survey for bright QSOs \citep{2000A&A...358...77W}, and the 6dF Galaxy Survey \citep{2009MNRAS.399..683J_6dfgs}, provided valuable quasar identifications but did not achieve the depth or uniformity seen in the north. Recently, efforts have been made in finding the brightest quasars in the southern hemisphere \citep[e.g.][]{2019ApJ...887..268C,2020ApJS..250...26B,2022MNRAS.510.2509C,2022MNRAS.511..572O,2023PASA...40...10O}, and systematic selections of quasar candidates in the Dark Energy Survey \citep{2023ApJS..264....9Y} and the KMTNet Synoptic Survey of Southern Sky \citep[][]{2024ApJS..275...46K}. 

In this paper, we extend the machine-learning purification of GDR3 QSO candidates to the southern equatorial hemisphere, which is only partly covered by PS1 and CatNorth. To do so we utilize optical to mid-infrared data from external datasets including the fourth data release of the SkyMapper Southern Survey \citep[SMSS DR4;][]{2024PASA...41...61O}, the second data release of the NOIRLab Source Catalog \citep[NSC DR2;][]{2021AJ....161..192N}, the VISTA \citep[Visible and Infrared Survey Telescope for Astronomy;][]{2006Msngr.126...41E} surveys, and CatWISE2020 in addition to \gaia DR3. We publish the results as the CatSouth quasar candidate catalog, and present a unified all-sky quasar candidate catalog (CatGlobe) by combining CatNorth and CatSouth.

This paper is organized as follows. In Section \ref{sec:data}, we introduce the data used in this work. Section \ref{sec:ml_selection} describes the machine-learning-based selection procedure of reliable quasar candidates. Section \ref{sec:redshift} describes the photometric redshift estimation and \gaia spectral redshift determination of the selected sample. Section \ref{sec:results} presents the contents and characteristics of the CatSouth and the all-sky CatGlobe quasar candidate catalogs. We summarize the paper in Section \ref{sec:conc}. 
% The Astronomical Data Query Language (ADQL) queries for selecting \gdr{3} stellar samples are included in Appendix \ref{adql:gaia}.
Throughout this paper we adopt a flat $\Lambda$CDM cosmology with $\Omega_{\Lambda}=0.7$, $\Omega_{\rm M}=0.3$, and $H_0=70\;\mathrm{km\;s^{-1}\;Mpc^{-1}}$. The $z$-band magnitude does not appear alone and will not be confused with the redshift symbol $z$. 

\section{Data} \label{sec:data}

The input data of this work is the \gdr{3} quasar candidate catalog (the \verb|qso_candidates| table) from \citet[][]{2023A&A...674A..41G} . We combine optical and infrared photometric data from \gdr{3}, SkyMapper DR4, CatWISE2020/AllWISE, and astrometric data from \gdr{3} to improve both purity and redshift estimation of the GDR3 QSO candidate catalog. We also retrieve samples of spectroscopically identified extragalactic objects from SDSS, and stellar samples from Gaia and additional catalogs to build well-defined training/validation sets. 

When querying the photometric data, all magnitudes are presented in their original system, i.e., Gaia and SkyMapper in the AB system, and WISE, 2MASS and VISTA in the Vega system. During the machine learning classification (Section \ref{sec:ml_selection}) and redshift estimation (Section \ref{sec:redshift}), we adopt all magnitudes in the AB system.

\subsection{Astrometric and Photometric Data} \label{sec:data-astrometry-photo}

\subsubsection{\gdr{3} Astrometric and Astrophysical Data}

\gdr{3} \citep{2023A&A...674A...1G} contains celestial positions, proper motions, parallaxes, and broadband photometry in the \textit{G}, {\gbp} (330--680~nm), and {\grp} (630--1050~nm) passbands for 1.8 billion sources at $G<21$ that have been present in the Early Third Data Release \citep[\gaia EDR3;][]{2021A&A...649A...1G}. Furthermore, the \gdr{3} catalog incorporates about 1 million mean spectra from the radial velocity spectrometer, about 220 million low-resolution blue and red prism photometer BP/RP mean spectra, variability results of 10 million sources across 24 variability types, astrophysical parameters for about 470 million source, and source class probabilities for 1,500 million sources, including stars, galaxies, and quasars. 

\subsubsection{CatWISE2020 and AllWISE Catalogs}
\label{sec:data-wise}

The Wide-field Infrared Survey Explorer \citep[WISE;][]{2010AJ....140.1868W} is a NASA Medium Class Explorer mission that conducted an imaging survey of the entire sky in the 3.4, 4.6, 12 and 22 $\mu$m mid-infrared bands (W1, W2, W3 and W4). The AllWISE source catalog \citep[][]{2019ipac.data...I1W} was built by combining data from the WISE cryogenic and NEOWISE \citep{2011ApJ...731...53M} post-cryogenic survey phases, providing positions, proper motions, four-band fluxes and flux variability statistics for over 747 million objects. 

The CatWISE2020 catalog \citep[][]{2021ApJS..253....8M,catwise_irsa551} consists of nearly 1.9 billion sources over the entire sky selected from the WISE cryogenic and NEOWISE post-cryogenic survey data at W1 and W2 bands collected from 2010 January 7 to 2018 December 13. CatWISE2020 has six times as many exposures spanning over 16 times as large a time baseline as the AllWISE catalog. The 5$\sigma$ limits for the CatWISE2020 Catalog in the Vega system are $\w{1}= 17.43$ mag and $\w{2}=16.47$ mag \citep[][]{2021ApJS..253....8M}. The 5$\sigma$ limiting magnitudes in the Vega system for the AllWISE catalog are 16.9, 16.0, 11.5, and 8.0 mag in W1, W2, W3 and W4, respectively\footnote{\url{https://wise2.ipac.caltech.edu/docs/release/allwise/expsup/sec2_3a.html}}. 

Because CatWISE2020 has deeper W1 and W2 data than AllWISE, we retrieve {{the point spread function (PSF) fitting magnitudes of W1 and W2}} from CatWISE2020 for \gdr{3} objects using pre-matched tables from NOIRLab datalab\footnote{\url{https://datalab.noirlab.edu/}} with a matching radius of $1\farcs5$. Through an outer join between this query and the AllWISE catalog, we obtain the W3 PSF magnitude from AllWISE as auxiliary data. 
% Entries in the AllWISE Source Catalog have been positionally cross-correlated with the Two Micron All Sky Survey \citep[2MASS;][]{2006AJ....131.1163S} Point Source Catalog (PSC) with a search radius of 3\arcsec and Extended Source Catalog (XSC).

We also set some constraints on the CatWISE2020 data. All sources should be: (i) not too bright to avoid possible saturation (\verb|w1mpro_pm>7 & w2mpro_pm>7|); (ii) significantly detected in W1 and W2 bands (\verb|w1snr_pm>5 & w2snr_pm>5|). 

\subsubsection{SkyMapper Southern Survey DR4}
The 1.3 m SkyMapper telescope at Siding Spring Observatory, Australia, has been conducting the SkyMapper Southern Survey \citep[SMSS;][]{2007PASA...24....1K,2018PASA...35...10W,2019PASA...36...33O} since 2014. The SkyMapper telescope has a 5.7 $\deg^2$ field-of-view, and the SMSS includes six optical filters: $u, v, g, r, i,$ and $z$ \citep{2011PASP..123..789B}. The fourth data release (DR4) of SMSS \citep{2024PASA...41...61O} covers a sky area of 26,000 $\deg^2$ from the South Celestial Pole to $\delta = +16 \degr$, with some fields of partial coverage reaching as far North as $\delta \sim 28 \degr$. The 10$\sigma$ depth in the $g$ band of SMSS DR4 is 20.5 mag (AB magnitude). 

We obtain SMSS DR4 PSF photometry for the \gdr{3} sources using pre-matched tables at NOIRLab datalab with a matching radius of $1\farcs5$. To select sources with good photometry, we require the $i$-band PSF magnitude $i>10$ (not saturated), and $i_\mathrm{err}<0.2171$ (equivalent to $\rm S/N>5$).

\subsubsection{VISTA Surveys}
VISTA \citep[Visible and Infrared Survey Telescope for Astronomy;][]{2006Msngr.126...41E} is a 4.1-m specialized wide-field survey telescope for the southern hemisphere. VISTA is equipped with a near-infrared camera VIRCAM \citep{2006SPIE.6269E..0XD} with a 1.65-degree diameter field of view, five available broadband filters at $Z,~Y,~J, ~H,~\ks$, and three narrow band filters at 0.98, 0.99, and 1.18 micron. VISTA has been conducting six large public surveys, covering different sky areas to different depths. These surveys include the VISTA Hemisphere Survey \citep[VHS;][]{2013Msngr.154...35M}, the VISTA Kilo-Degree Infrared Galaxy Survey \citep[VIKING;][]{2013Msngr.154...32E}, the VISTA Magellanic Survey \citep[VMC;][]{2011Msngr.144...25C,2011A&A...527A.116C}, the VISTA Variables in the Via Lactea survey \citep[VVV;][]{2010NewA...15..433M}, the VISTA Deep Extragalactic Observations Survey \citep[VIDEO;][]{2013Msngr.154...26J,2013MNRAS.428.1281J}, and UltraVISTA \citep{2012A&A...544A.156M}. 

The VHS is imaging the entire southern hemisphere of the sky, except the areas covered by the VIKING, VMC, and VVV surveys. Therefore, combining data from VHS and other VISTA surveys will improve the sky coverage of near-infrared photometric data in the southern hemisphere. We perform outer joins between our samples and the VHS, VIKING, VMC, and VVV surveys, each from a specific data release and source: VHS DR5 (ESO version 3) from the NOIRLab Data Lab\footnote{\url{https://datalab.noirlab.edu/vhsdr5.php}}; VVV DR4.2 from VizieR\footnote{\url{https://cdsarc.cds.unistra.fr/viz-bin/w/VizieR?-source=II/376}} through the VizieR TAP service\footnote{\url{http://tapvizier.cds.unistra.fr/TAPVizieR/tap}}; VIKING DR4 via the VSA TAP service\footnote{\url{http://tap.roe.ac.uk/vsa}}; and VMC DR6 (ESO version 5) through the ESO TAP service\footnote{\url{https://archive.eso.org/tap_cat}}. A matching radius of 1\farcs5 is used when joining our samples with the VHS, VIKING, and VMC catalogs, and a matching radius of 1\farcs0 is used when joining our samples with the VVV catalog to avoid mismatches in the crowded Galactic plane. 

{{We use the default point source aperture photometry (\texttt{APERMAG3}) from VISTA, which has been aperture corrected and is suitable for point sources even in crowded fields \citep[][]{2018MNRAS.474.5459G}. }}
To increase the query efficiency and ensure the data quality of the extremely large VIKING, VMC, and VVV catalogs, we set the following constraints when querying the databases:
\begin{verbatim}
    (YAPERMAG3 BETWEEN 11 AND 22) 
    AND (JAPERMAG3 BETWEEN 11 AND 22) 
    AND (KSAPERMAG3 BETWEEN 11 AND 22) 
    AND MERGEDCLASS NOT IN (0, -9)
    AND JAPERMAG3ERR < 0.2171
    AND YPPERRBITS<256
    AND JPPERRBITS<256
    AND KSPPERRBITS<256.
\end{verbatim}

% An example of such data combination is the VISTA EXtension to Auxiliary Surveys catalog \citep[VEXAS;][]{2019A&A...630A.146S}, which combines VHS and VIKING and 

\subsubsection{NOIRLab Source Catalog DR2}
The second data release (DR2) of the NOIRLab Source Catalog \citep[NSC;][]{2018AJ....156..131N,2021AJ....161..192N} is a catalog of over 3.9 billion sources from public imaging data in NOIRLab's Astro Data Archive\footnote{\url{https://astroarchive.noirlab.edu/}}. Most of the images used in NSC DR2 are taken with CTIO-4 m Blanco + DECam (340,952 exposures). In addition, there are 41,561 exposures from KPNO 4-m Mayall + Mosaic3 \citep[the majority from the Mayall $z$-band Legacy Survey, MzLS;][]{2016SPIE.9908E..2CD} and 29,603 exposures from the Steward Observatory Bok-2.3 m + 90Prime \citep[from the Beijing-Arizona Sky Survey, BASS;][]{2017PASP..129f4101Z,2018ApJS..237...37Z,2019ApJS..245....4Z}. A large fraction of the images are data obtained by the Dark Energy Survey \citep[DES;][]{2017ApJ...848L..13A} and the Legacy Survey imaging projects \citep{2019AJ....157..168D}.
% 412,116 public images from CTIO-4m+DECam, KPNO-4m+Mosaic3, and the Bok-2.3+90Prime have been used to generate the NSC DR2. 
NSC DR2 includes photometry in 7 bands, namely $u$, $g$, $r$, $i$, $z$, $Y$, and \textit{VR}, with different sky coverages and depths. 

We utilize data of NSC DR2 $griz$ bands to impute missing values of SMSS $griz$ bands. We do not apply any transformation to the NSC DR2 $griz$ photometry when imputing the missing values because the magnitude offsets between the two datasets vary across different sky regions \citep[see Figure 19 of][]{2024PASA...41...61O} and are sensitive to extinction. Nevertheless, such missing data imputation is still very helpful for training and applying machine learning models, because for a list of sources, valid values from a similar NSC band are more informative and discriminative than a single mean or median value of the whole sample.

\subsubsection{2MASS}
The Two Micron All Sky Survey \citep[2MASS;][]{2003ipac.data...I2S,2006AJ....131.1163S} has uniformly scanned the entire sky in $J$ (1.25 microns), $H$ (1.65 microns), and \ks (2.17 microns) bands using two 1.3-m telescopes, one at Mt. Hopkins, Arizona, and one at CTIO, Chile. Because the AllWISE Source Catalog that we use (Section \ref{sec:data-wise}) includes the closest entries from the 2MASS Point Source Catalog (PSC) within 3\arcsec~from the AllWISE positions\footnote{\url{https://wise2.ipac.caltech.edu/docs/release/allwise/expsup/sec2_2.html}}, we adopt the 2MASS photometry from AllWISE directly. We use the 2MASS $JH\ks$ magnitudes to impute missing values of VISTA photometry by converting 2MASS magnitudes to VISTA ones. To do so, the offsets between the median values of 2MASS bands and those of the VISTA bands are calculated and subtracted from 2MASS magnitudes for each of the training/validation samples (stars, galaxies, quasars), and the test sample (GDR3 QSO candidates). 

\subsubsection{Relative Extinction Coefficients}

For each photometric band we use in this work, we compute its relative extinction coefficient $R_{\lambda\mathrm{p}}$, defined as

\begin{equation}
    R_{\lambda\mathrm{p}} = \frac{A_{\lambda\mathrm{p}}}{A_V} \times R_V,
\end{equation}

\noindent where $A_{\lambda\mathrm{p}}$ and $A_V$ are the extinctions in the band with pivot wavelength $\lambda_{\mathrm{p}}$ and $V$ band, and $R_V=A_V/E(B-V)$. The pivot wavelength $\lambda_{\mathrm{p}}$ \citep{1986HiA.....7..833K} is calculated as 

\begin{equation}
\lambda_\mathrm{p} = \sqrt{\frac{\int_\lambda T(\lambda)\,\lambda\,d\lambda}{\int_\lambda T(\lambda)\,d\lambda /\lambda}},
\end{equation}

\noindent where $T(\lambda)$ is the throughput (transmission curve) of the filter as a function of the wavelength.

Using the definition of passbands from \citet{2010AJ....140.1868W,2011PASP..123..789B,2018MNRAS.474.5459G,2021A&A...649A...3R}, and the optical to mid-IR extinction law from \citet{wang2019optical} assuming $R_{V}=3.1$, we calculate the extinction coefficients as listed in Table \ref{tab:ext_coef}. The coefficients are used for Galactic extinction correction during the photometric redshift estimation procedure.

\begin{deluxetable}{cccc}
\tablecaption{Relative extinction coefficients of passbands used in this work ($R_V=3.1$). \label{tab:ext_coef}}
\tablehead{\colhead{Band} & \colhead{$\lambda_{\mathrm{p}}$ (\AA)} & \colhead{$A_{\lambda\mathrm{p}}/A_V$} & \colhead{$R_{\lambda\mathrm{p}}$}}
\startdata
$g$ & 5075.19 & 1.1100 & 3.4411 \\
$r$ & 6138.44 & 0.8535 & 2.6459 \\
$i$ & 7767.98 & 0.5851 & 1.8139 \\
$z$ & 9145.99 & 0.4382 & 1.3585 \\
\textit{G} & 6217.59 & 0.8373 & 2.5957 \\
\gbp & 5109.71 & 1.1005 & 3.4116 \\
\grp & 7769.02 & 0.5850 & 1.8135 \\
$Y$ & 10210.71 & 0.3565 & 1.1051 \\
$J$ & 12524.83 & 0.2336 & 0.7240 \\
$H$ & 16432.45 & 0.1331 & 0.4127 \\
\ks & 21521.52 & 0.0762 & 0.2361 \\
W1 & 33682.21 & 0.0301 & 0.0934 \\
W2 & 46179.06 & 0.0157 & 0.0486 \\
W3 & 120718.09 & 0.0021 & 0.0067 
\enddata
\end{deluxetable}

\subsection{Stellar Sample} \label{sec:stellar_samples}

A robust stellar sample is essential for constructing the training set of our machine learning classifier and for minimizing contaminant misclassifications. In CatSouth we adopt a strategy similar to that in \citet{2024ApJS..271...54F}: we select stars from \gdr{3} within the SMSS DR4 footprint and complement this primary sample with additional very low-mass stars (VLMS), white dwarfs (WD), and carbon stars to form the combined master stellar sample. 

\subsubsection{O-to-M Type Stars from \gdr{3}}\label{sec:om_stars}
We first extract a representative sample of O-to-M type stars from \gdr{3} following the methods in \citet{2024ApJS..271...54F}. In particular, we purify the original \gdr{3} OBA gold sample by excluding sources with tangential velocity ($v_{\mathrm{tan}}$) higher than $180\;\mathrm{km\;s^{-1}}$ as suggested by \citet{2023A&A...674A..39G}, and increase the completeness of the FGKM sample using a less strict selection than the one adopted by \citet{2023A&A...674A..39G}. Only sources with declinations satisfying $\delta<16\degr$ are selected to effectively match the SMSS DR4 footprint. The detailed ADQL queries used for the Gaia selections of O-to-M type stars are provided in Appendix~\ref{adql:gaia}. Finally, these stars are crossmatched with other catalogs in Section \ref{sec:data-astrometry-photo} with the corresponding photometric quality constraints.

\subsubsection{Additional VLMS, White Dwarfs, and Carbon Stars}\label{sec:extra_stars}
To account for atypical/under-representative stars that may contaminate the quasar selection, we supplement the O-to-M type sample with additional VLMS, WDs, and carbon stars compiled from the literature. Major sources for these additional samples include \citet{2021ApJS..253...45L}, \citet{2011AJ....141...97W}, and \citet{2001BaltA..10....1A}, and we refer to Table~1 in \citet{2024ApJS..271...54F} for a complete list of the catalogs. These extra stars are crossmatched with catalogs in Section \ref{sec:data-astrometry-photo}, and are merged with the O-to-M type stars to form the master stellar sample. This combined stellar sample contains more than 1.84 million sources satisfying the photometric quality constraints in Section \ref{sec:data-astrometry-photo}, providing a diverse and comprehensive training set for our classification model.

\subsection{Quasar Sample} \label{subsec:data-qso}
We combine known quasars from the Million Quasars catalog \citep[Milliquas v8;][]{2023OJAp....6E..49F}, and highly reliable quasar candidates from the CatNorth quasar candidate catalog \citep{2024ApJS..271...54F} to build the quasar sample for training the machine classification model. 

Milliquas v8 is a compilation of quasars and quasar candidates from the literature up to 2023 June 30, which includes 907,144 type 1 QSOs and AGNs, 66,026 high-confidence (pQSO=99\%) photometric quasar candidates, 2814 BL Lac objects, and 45,816 type 2 objects. We select the sub-sample of Milliquas by requiring that the sources are: (i) located at $\delta < 16 \degr$, and (ii) spectroscopically identified type 1 quasars with valid redshifts (labeled as ``Q'' in the ``TYPE'' column of Milliquas, and $z>0$). 

Among the 290,294 selected quasars, about 75\% are identified by SDSS \citep[e.g.][]{2020ApJS..249....3A,2020ApJS..250....8L,2023ApJS..267...44A}; other major sources come from the DESI EDR \citep{2024AJ....168...58D}, the 2dF QSO Redshift Survey \citep[2QZ;][]{2004MNRAS.349.1397C}, the LAMOST quasar survey \citep[][]{2016AJ....151...24A,2018AJ....155..189D,2019ApJS..240....6Y,2023ApJS..265...25J}, and the 2dF-SDSS LRG and QSO survey \citep[2SLAQ;][]{2009MNRAS.399.1755C}. 

CatNorth is a quasar candidate catalog built upon the GDR3 QSO candidate catalog using photometric data from Gaia, PS1, CatWISE, and low-resolution spectral information from Gaia. To complement the Milliquas sample, we select a highly reliable sub-sample of CatNorth in the SMSS footprint using the following criteria: 
\begin{enumerate}
    \item $p_{\mathrm{QSO\_mean}} > 0.95$,
    \item $\delta < 16 \degr$,
    \item $|z_{\mathrm{ph}}-z_{\gaia}|/(1+z_{\mathrm{ph}})<0.02$, 
\end{enumerate}
where $p_{\mathrm{QSO\_mean}}$ is the machine-learning predicted probability of a source being a quasar, $z_{\mathrm{ph}}$ is the photometric redshift given by CatNorth, and $z_{\gaia}$ is the redshift in the \gdr{3} quasar candidate catalog (\verb|redshift_qsoc|). The selection yields 323,626 high-confidence quasar candidates. 

After crossmatched with \gdr{3}, SMSS DR4, and CatWISE2020, and filtered with the corresponding quality constraints, 105,413 sources remain in the Milliquas v8 subsample, and 276,814 sources remain in the CatNorth subsample. Combining the two subsets gives the final quasar sample containing 282,322 unique sources.

\subsection{Galaxy Sample}
The galaxy sample is built upon the combination of the spectroscopic galaxy catalogs of the Seventeenth Data Release of the Sloan Digital Sky Surveys \citep[SDSS DR17;][]{2022ApJS..259...35A}, the 2dF Galaxy Redshift Survey \citep[2dFGRS;][]{2001MNRAS.328.1039C_2dfgrs}, the 6dF Galaxy Survey \citep[6DFGS;][]{2009MNRAS.399..683J_6dfgs}, and the 2MASS Redshift Survey \citep[2MRS;][]{2012ApJS..199...26H_2mrs}. In particular, the SDSS DR17 galaxy sample is selected from the SpecObj table\footnote{\url{https://data.sdss.org/sas/dr17/sdss/spectro/redux/specObj-dr17.fits}} using the following criteria:

\begin{enumerate}
    \item The objects are spectroscopically classified as galaxies without broad emission lines ($\sigma_{\mathrm{line}}>200\ \mathrm{km\ s^{-1}}$) detected at the 5-sigma level: \verb|CLASS == 'GALAXY' AND SUBCLASS|\\ \verb|NOT LIKE 'BROADLINE'|.
    \item The spectra are primary detections with good observational conditions and high S/N, and no issues are found in fitting the redshifts: \verb|SPECPRIMARY == 1 AND| \verb|PLATEQUALITY == 'good' AND|\\ \verb|SN_MEDIAN_ALL > 5 AND ZWARNING == 0|.
\end{enumerate}

For the other three catalogs (2dFGRS, 6DFGS, and 2MRS), we select galaxies that are at $\delta<16\degr$ with valid redshifts ($z>0$), and not classified as quasars / broad-line AGN by any of the following catalogs:

\begin{enumerate}
    \item Milliquas v8.
    \item A Uniformly Selected, All-sky, Optical AGN Catalog by \citet{2019ApJ...872..134Z}.
    \item An extensive list of broad-line AGN in the 6dF Galaxy Survey \citep{2025MNRAS.536.3611H}.
\end{enumerate}

We merge all galaxies satisfying the photometric quality constraints in Section \ref{sec:data-astrometry-photo} into one sample, which contains 382,303 unique sources using an inner match radius of 1\farcs5. Finally, the labeled stars, quasars, and galaxies are combined as a unified training/validation set with more than 2.5 million sources.

\section{Machine-learning Selection of Quasar Candidates} \label{sec:ml_selection}
% \subsection{Feature Selection and Characterization}
\subsection{XGBoost Classification of Stars, Galaxies, and Quasars}

We use XGBoost \citep[][]{chen2016xgboost}, a gradient boosting decision tree algorithm to train the machine learning classifier using the training/validation set, and reclassify the input \gdr{3} quasar candidates as quasars, stars, and galaxies. Gradient boosting tree algorithms including XGBoost have shown exceptional performance in astronomical studies, especially those on quasar candidate selections and photometric redshifts \citep[e.g.][]{2019MNRAS.485.4539J,fu2021gpq1,fu2022gpq2,2024ApJS..271...54F,2022MNRAS.509.2289L,2022A&A...668A..99H,2024PASJ...76..653K,2024ApJS..275...19Y}.

We follow the feature selection strategy of \citet[Section 3]{2024ApJS..271...54F}, which combines two types of features:
(1) broadband colors that capture differences in spectral energy distributions (SEDs) among quasars, stars, and galaxies, and
(2) morphological indicators that help distinguish point-like sources (such as stars and quasars) from extended sources (such as galaxies). Similarly, we construct a multi-dimensional feature set for training the classification model in this work using multi-wavelength data described in Section \ref{sec:data}. The color and magnitude features include \w{1} (the W1 magnitude), and a set of color indices, $g-r$, $r-i$, $i-z$, $i-J$, $J-\ks$, $i-\w{1}$, $z-\w{1}$, $J-\w{1}$, $\w{1}-\w{2}$, $\w{2}-\w{3}$, $G_{\mathrm{BP}}-G_{\mathrm{RP}}$, $G_{\mathrm{BP}}-G$, and $G-G_{\mathrm{RP}}$, all in AB magnitudes. We do not apply extinction corrections to the magnitudes and colors because such corrections depend on source types and distances, which are not known for the sources in the application (test) set. We also compute two morphological indicators, $\log(\chi^2_{\mathrm{PSF}})$ from SMSS DR4, and $\log(1+C^*)$ from \gdr{3}, which are discussed later in this section. Together, 16 features are used to train the XGBoost classifier.

The magnitude and colors we choose capture key differences in the SEDs of quasars versus stars and galaxies. In particular, the mid-infrared color $\w{1}-\w{2}$ has been proven effective in selecting AGNs due to their power-law SEDs and hot dust emission \citep[e.g.,][]{2012ApJ...753...30S,2012AJ....144...49W,2018ApJS..234...23A}. The near infrared color $J-\ks$ is also powerful in separating quasars at various redshifts from stars \citep[e.g.][]{2008MNRAS.386.1605M,2010MNRAS.406.1583W,2017ApJ...851...13S}. The combination of the optical color indices ($g-r$, $r-i$, $i-z$) can help reduce the overlap between quasars and the stellar loci on two-dimensional color-color diagrams \citep{fu2021gpq1}.

To complement the color and magnitude information, we include two morphological features sensitive to source extent. The first parameter is $\chi^2_{\mathrm{PSF}}$, defined as the maximum chi-squared value from the PSF photometry across the available SMSS DR4 bands \citep[see Sec. 6.7.4 of][]{2024PASA...41...61O}. Point sources typically yield $\chi^2_{\mathrm{PSF}}$ values between 1 and 3, while extended objects have substantially larger values, e.g., up to a few thousand for galaxies. The second morphological feature is the corrected BP/RP flux excess factor, $C^*$, from Gaia DR3 \citep{2021A&A...649A...3R}\footnote{\url{https://github.com/agabrown/gaiaedr3-flux-excess-correction}}. The original BP/RP flux excess factor, $C$, is the ratio of the sum of the integrated BP and RP fluxes to the flux in the \textit{G} band: $C=(I_{\mathrm{BP}}+I_{\mathrm{RP}})/I_{G}$. Because the detection windows (apertures) of BP and RP bands are wider than that of the \textit{G} band, extended sources tend to have larger flux excess factors than the point sources do \citep[see e.g.][]{2020ApJS..250...17L}. The corrected BP/RP flux excess factor $C^*$ removes the color dependence present in the original $C$, enabling more robust characterization of source extent \citep{2021A&A...649A...3R}. We use the logarithmically transformed morphological features, $\log(\chi^2_{\mathrm{PSF}})$ and $\log(1+C^*)$, to compress the dynamic ranges of the original values and mitigate the influence of extreme outliers. 

\begin{figure*}[ht]
    \centering
    \includegraphics[width=1\textwidth]{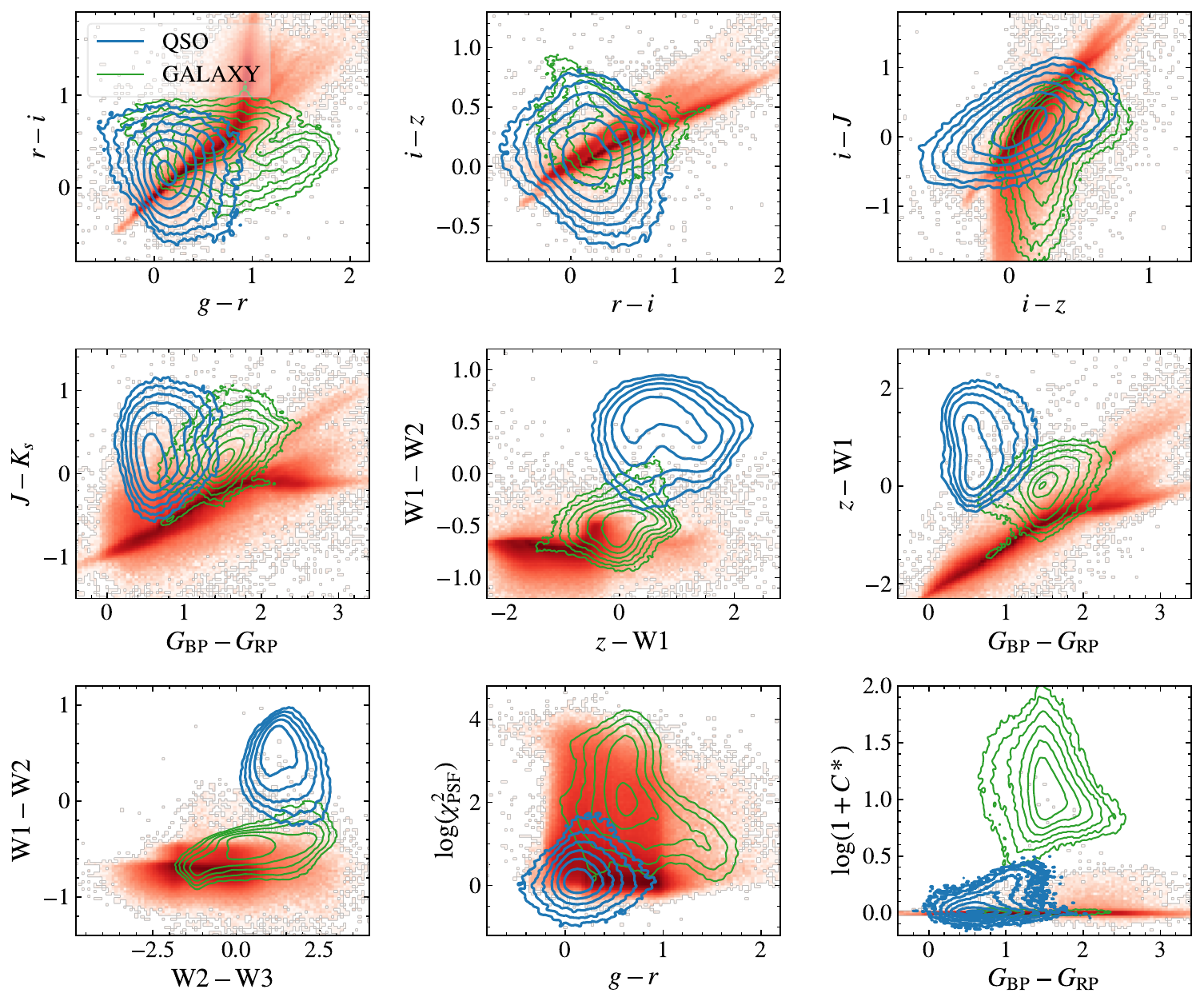}
    \caption{Two-dimensional feature representations (color-color and morphology-color diagrams) of sources in the training sample. Quasars are shown as blue contours, galaxies as green contours, and stars as red-shaded density plots. All magnitudes are in the AB system and not dereddened.}
    \label{fig:ccd_trainval}
\end{figure*}

Figure \ref{fig:ccd_trainval} shows the two-dimensional representations of some selected combinations of features of the training/validation set. {{Stars form narrow, concentrated loci on color-color diagrams that blend with quasars in the optical. The stellar loci become more distinctly separated from quasars in the infrared bands.}} Quasars from the training set produce unimodal probability distributions with smooth boundaries on the color-color and morphology-color diagrams, indicating no contamination from stars. However, the outermost contours of galaxies from the training set show spikes that overlap with stellar loci, indicating minor contributions from stars. We do not purify the galaxy sample further because our primary goal is to optimize quasar selection; since both galaxies and stars are treated as contaminants, the small level of stellar contamination in the galaxy sample does not impact the performance of the classifier in identifying quasars. 

We use a few metrics to evaluate the model performance. For binary classification problems, with true positive denoted as TP, true negative as TN, false positive as FP, and false negative as FN, the metrics are defined as: 

\begin{gather}
    % \rm balanced\ accuracy = \frac{1}{2}\left( \frac{TP}{TP + FN} + \frac{TN}{TN + FP}\right )\\
    \rm precision = \frac{TP}{TP+FP}\\
    \rm recall = \frac{TP}{TP+FN}\\
    F_{1} = \rm 2\times \frac { precision \times recall}{precision + recall} .
\end{gather}

\noindent In the case of a multiclass problem, the classification task is treated as a collection of binary classification problems, one for each class. The metrics above can be calculated for each binary classification problem (each class). The metrics of the multiclass problem are the average metrics of all classes.

Hyperparameter optimization is performed with Optuna \citep{akiba2019optuna} via five-fold cross validation, minimizing the multi-class log loss as the objective function among 500 trials. We refer to \citet{2024ApJS..271...54F} for a detailed hyperparameter tuning procedure. After obtaining the optimal hyperparameters, the whole input data is split into a training set and a validation set according to a $4:1$ ratio, and the optimal classifier is trained on the entire training set.

\begin{figure}[ht]
    \centering
    \includegraphics[width=0.49\textwidth]{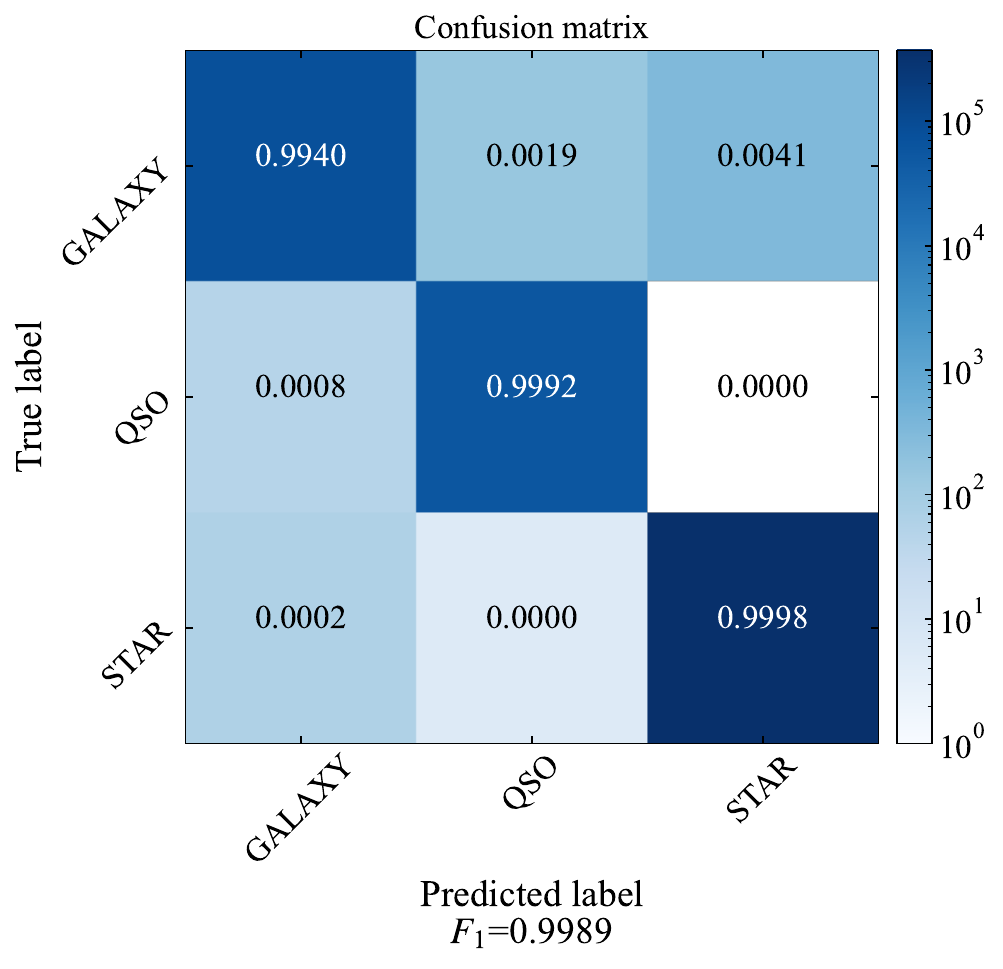}
    \caption{
    % Normalized confusion matrix of the XGBoost classifier computed with the validation set. The matrix is color-coded with number of sources of each element. The diagonal elements represent the number of objects for which the predicted label equals to the true label, and the normalized fractions shown in the diagonal elements represent the recall rates (completeness) of the three classes.
    Normalized confusion matrix of the XGBoost classifier computed on the validation set. The matrix is color-coded by the number of sources in each cell. Diagonal entries show the fraction of correctly classified objects (i.e., recall or completeness) for each class, while off-diagonal entries indicate the misclassification rates.
    }
    \label{fig:xgb_clf_cm}
\end{figure}

The normalized confusion matrix for the three-class problem, calculated on the validation set and shown in Figure \ref{fig:xgb_clf_cm}, demonstrates the outstanding performance of the trained XGBoost classifier. For example, 99.40\% of galaxies, 99.92\% of quasars, and 99.98\% of stars are correctly identified, while the off-diagonal elements indicate very low misclassification rates. The actual fraction of galaxies misclassified as stars is expected to be even lower than 0.41\% shown in the confusion matrix, because the training/validation sample of galaxies is contaminated by stars. The good performance is also reflected in the overall $F_1$ score of 0.9989, which confirms that both precision and recall are extremely high for all classes. 
% Such results prove that the model reliably distinguishes between galaxies, quasars, and stars.

We apply the optimal XGBoost classifier to the input test sample of \gdr{3} quasar candidates in this work, and assign probability estimates ($p_{\mathrm{QSO}}$, $p_{\mathrm{star}}$, $p_{\mathrm{galaxy}}$) for each source. By default, a source is labeled as the class that receives the highest probability.

\subsection{Additional filtering to remove contaminants}
\label{sec:pm_filter}

To further reduce stellar contamination in our quasar candidate sample, we apply a probabilistic proper motion cut based on the likelihood that a source has zero proper motion. Following the method described in \citet{fu2021gpq1,2024ApJS..271...54F}, the probability density function of zero proper motion, $f_{\mathrm{PM0}}$, is defined as

\begin{multline}
        f_{\mathrm{PM0}} = \frac{1}{2\pi \sigma_x \sigma_y\sqrt{1-\rho^2}} \times \\
    \mathrm{exp}\left\{-\frac{1}{2(1-\rho^2)}\left[ \left(\frac{x}{\sigma_{x}}\right)^{2} - \frac{2\rho xy}{\sigma_{x}\sigma_{y}} + \left(\frac{y}{\sigma_{y}}\right)^{2} \right] \right\},
\end{multline}

\noindent where $x$ is the proper motion in right ascension (\texttt{pmra}), $y$ is the proper motion in declination (\texttt{pmdec}), $\rho$ is the correlation coefficient between $x$ and $y$ (\texttt{pmra\_pmdec\_corr}), and $\sigma_x$ and $\sigma_y$ are the proper-motion uncertainties. In practice, for a given level of uncertainty, sources with smaller proper motions yield higher values of $f_{\mathrm{PM0}}$. To facilitate comparison across different samples, we work with the logarithm of this quantity, $\log (f_{\mathrm{PM0}})$. As demonstrated in \citet{2024ApJS..271...54F}, a threshold of $\log (f_{\mathrm{PM0}}) \geq -4$ effectively excludes over 99.9\% of stellar contaminants while retaining more than 99.8\% of the quasar sample. 

% This threshold helps mitigate the influence of stars with spurious proper motion measurements, thereby ensuring that our quasar selection remains robust. For further details on the derivation and performance of this proper motion-based cut, we refer the reader to \citet{fu2021gpq1} and our CatNorth paper.

% In order to remove stellar contaminants such as white dwarfs, M/L/T dwarfs, YSOs, and AGB stars from quasar candidates, we apply an additional cut based on \gaia\ proper motion, because the proper motion distribution of quasars is different from that of Milky Way stars. Although quasars should have negligible transverse motions, non-zero proper motions of them are measured by \gaia\ due to various effects, such as photocenter variability of quasars \citep[see][and references therein]{2016AAP...589A..71B}, and double/multiple sources \citep{2022ApJ...933...28M}. In addition, proper motions with large uncertainties are not reliable. Therefore we need a probabilistic cut instead of a cut on the total proper motion. In \citet{fu2021gpq1}, we defined the probability density of zero proper motion ($f_{\mathrm{PM0}}$) of a source, based on the bivariate normal distribution of proper motion measurements of the source as:

% \begin{multline}
%         f_{\mathrm{PM0}} = \frac{1}{2\pi \sigma_x \sigma_y\sqrt{1-\rho^2}} \times \\
%     \mathrm{exp}\left\{-\frac{1}{2(1-\rho^2)}\left[ \left(\frac{x}{\sigma_{x}}\right)^{2} - \frac{2\rho xy}{\sigma_{x}\sigma_{y}} + \left(\frac{y}{\sigma_{y}}\right)^{2} \right] \right\},
% \end{multline}

To account for the higher source density and increased contamination in the regions around the Large and Small Magellanic Clouds (LMC and SMC), we define two circular regions centered on these objects. Specifically, we construct a $10\degr$ radius region centered at LMC ($\alpha = 80.8942\degr$, $\delta = -69.7561\degr$) and a $5\degr$ radius region centered at SMC ($\alpha = 13.1583\degr$, $\delta = -72.8003\degr$) using multi-order coverage (MOC) maps generated with \texttt{MOCPy} \citep{matthieu_baumann_2024_14205461}. We then determine which sources in our catalog fall within these regions by verifying whether their coordinates are contained in the corresponding MOCs.

Based on the spatial locations, we apply distinct selection criteria to select reliable quasar candidates. For sources outside the LMC/SMC regions, where contamination is lower, we require $\log f_{\mathrm{PM0}} > -4$ and that the predicted class is ``QSO" (i.e. $p_\mathrm{QSO}>p_\mathrm{star}~\&~p_\mathrm{QSO}>p_\mathrm{galaxy} $). In contrast, for sources within the LMC or SMC regions, which are known to exhibit higher stellar crowding and potential misclassification, we enforce a stricter threshold by requiring $\log f_{\mathrm{PM0}} > -1$ and $p_\mathrm{QSO} > 0.9$. This dual-threshold strategy ensures a more reliable selection of quasar candidates across different sky regions, maintaining high purity in crowded fields while preserving completeness in less contaminated areas. In total, 921,528 sources out of 1,174,509 input GDR3 QSO candidates are selected as reliable quasar candidates.

\section{Redshift estimation for quasar candidates} \label{sec:redshift}

We estimate photometric redshifts for all quasar candidates selected from above, and spectroscopic redshifts for quasar candidates with available Gaia BP/RP spectra. For both regression models, we adopt the root mean square error (RMSE), the normalized median absolute deviation of errors ($\sigma_{\mathrm{NMAD}}$), and the catastrophic outlier fraction ($f_{\mathrm{c}}$)  as evaluation metrics for the redshift estimation in the training/validation sets. These metrics are defined as follows:
\begin{gather}
    \mathrm{RMSE} = \sqrt{ {\frac {1}{n}}\sum _{i=1}^{n}\left(z_{i}-{\hat {z_{i}}}\right)^{2} } ,\\
    \sigma_{\mathrm{NMAD}} = 1.48 \times \mathrm{median} \left(\left| \frac{\Delta z- \mathrm{median}(\Delta z)}{1+z} \right|\right) , \\
    f_{\mathrm{c}} = \frac{1}{n} \times \mathrm{count}\left(\left| \frac{\Delta z}{1+z} \right| > 0.15\right),
\end{gather}

\noindent where $z$ is the true redshift, $\hat{z}$ is the predicted redshift, $\Delta z = z-\hat{z}$, and $n$ is the total number of sources. The RMSE is widely used in regression analysis to quantify the difference between the true and predicted values. The $\sigma_{\mathrm{NMAD}}$ measures the statistical dispersion of the normalized errors $\Delta z^{\prime} = \Delta z /(1+z)$ \citep{2006A&A...457..841I,2008ApJ...686.1503B}. 
% while less sensitive to outliers than the original standard deviation.  
The $f_{\mathrm{c}}$ represents the percentage of objects for which the redshift estimate deviates significantly from the true redshift. 

% While photometric redshift estimation relies on a good training sample with known spectroscopic redshifts,  

\subsection{BP/RP Spectroscopic Redshift Determination}\label{sec:specz}

In \citet{2024ApJS..271...54F}, we have trained a convolutional neural network (CNN) regression model (RegNet) for quasar redshift determination using Gaia BP/RP spectra of quasars with known redshifts, which achieved $\mathrm{RMSE}=0.1427$, $\sigma_{\mathrm{NMAD}}=0.0304$, and $f_{\mathrm{c}}=2.46\%$ on the validation set. Because Gaia BP/RP spectra do not rely on external photometric surveys, the spectroscopic redshifts of CatSouth sources with Gaia BP/RP spectra can be easily determined using the pre-trained RegNet model.

The BP/RP spectra of CatSouth sources are retrieved via the \verb|astroquery.gaia| module, then calibrated and resampled over the wavelength range $\mathrm{[4000\AA, 10000\AA)}$ in 20\,\AA\ intervals using the GaiaXPy package \citep{daniela_ruz_mieres_2023_7566303}. Each spectrum is represented by 300 data points and normalized to the $[0,1]$ range before inference with RegNet. The redshift estimates from RegNet are denoted as $z_{\rm xp\_nn}$. 

% Because $z_{\rm xp\_nn}$ does not rely on external photometric data such as SMSS DR4 and VISTA catalogs, this 

% We provide these spectroscopic redshifts as an independent 
% an integral component in constructing the photometric redshift training sample (see next subsection). In particular, CatSouth sources with Gaia BP/RP spectra for which $z_{\rm xp\_nn}$ is in excellent agreement with the Gaia QSOC redshift ($z_{\rm qsoc}$) are flagged as high-quality; their combined redshift, defined as 
% \[
% z_{\rm cat} = \frac{z_{\rm xp\_nn} + z_{\rm qsoc}}{2},
% \]
% is then used in the subsequent photometric redshift estimation.

\subsection{Ensemble Photometric Redshift Estimation with XGBoost, TabNet, and FT-Transformer} \label{sec:photoz}

Photometric redshift estimation with machine learning algorithms requires a training sample with good data quality in both feature columns and spectroscopic redshift. Starting from the 105,413 quasars from Milliquas that satisfy the photometric quality constraints in Section \ref{subsec:data-qso}), we perform the following procedures to build a subsample with accurate and precise spectroscopic redshifts:

\begin{enumerate}
    \item Because Milliquas only preserves three digits for the redshift (column ``Z''), we replace the Milliquas redshift values ($z_{\mathrm{MQ}}$) with the ones with higher precision from SDSS DR16Q \citep[$z_{\mathrm{sys}}$ from][]{2022ApJS..263...42W}, SDSS DR18 \citep{2023ApJS..267...44A}, and DESI EDR \citep{2024AJ....168...58D} when available. The spectroscopic redshift values from other origins are kept as they are. The newly adopted redshift is denoted as $z_{\mathrm{cat}}$.
    \item For DR16Q sources in the subset, we keep those with robust redshift estimates by requiring: $z_{\mathrm{sys}}>0$, 
    $z_{\mathrm{sys\_error}}\neq-1$, $z_{\mathrm{sys\_error}}\neq-2$, $z_{\mathrm{sys\_error}}/(1+z_{\mathrm{sys}})<0.002$ and $|z_{\mathrm{sys}}-z_{\mathrm{DR16Q}}|/(1+z_{\mathrm{sys}})<0.002$, where $z_{\mathrm{DR16Q}}$ is the final redshift from \citet{2020ApJS..250....8L}.
    \item Milliquas has made corrections for a small fraction of quasars with wrong redshifts from SDSS or DESI. Therefore we use $|z_{\mathrm{MQ}}-z_{\mathrm{cat}}|/(1+z_{\mathrm{cat}})<0.002$ to select quasars with good redshifts in both Milliquas and original catalogs.
\end{enumerate}

The resulted spectroscopic Milliquas sub-sample contains 96,181 sources. To complement this spectroscopic sample, we select additional quasars from CatSouth that have CNN-derived redshifts in very close agreement with the Gaia redshifts by requiring $|z_{\mathrm{xp\_nn}}-z_{\mathrm{Gaia}}|/(1+z_{\mathrm{Gaia}})<0.01$. For these additional sources, we compute the target redshift $z_{\rm cat}$ by averaging the CNN-derived $z_{\rm xp\_nn}$ and the original Gaia redshift ($z_{\rm Gaia}$), i.e.,

\begin{equation}
    z_{\rm cat} = \frac{z_{\rm xp\_nn} + z_{\rm Gaia}}{2}.
\end{equation}

\noindent The combined training sample for the photometric redshift estimation contains 108,885 unique sources. 

We choose a total of 18 features for training and deploying the machine learning models: W1, $g-r$, $r-i$, $i-z$, $i-J$, $J-\ks$, $i-$W1, $z-$W1, $J-$W1, W1$-$W2, W2$-$W3, $G_{\mathrm{BP}}-G_{\mathrm{RP}}$, $G_{\mathrm{BP}}-G$, $G-G_{\mathrm{RP}}$, $\log(\chi^2_{\mathrm{PSF}})$, $\log(1+C^*)$, $\log(1+z_{\mathrm{low}})$, and $\log(1+z_{\mathrm{up}})$. Here, $z_{\mathrm{low}}$ (\texttt{redshift\_qsoc\_lower}) and $z_{\mathrm{up}}$ (\texttt{redshift\_qsoc\_upper}) are the lower and upper confidence intervals of $z_{\mathrm{Gaia}}$ taken at 0.15866 and 0.84134 quantiles, respectively. All magnitudes are in the AB system, and corrected for Galactic dust extinction using the \citet{2016A&A...596A.109P} dust map and the extinction coefficients in Table \ref{tab:ext_coef}. Missing values are imputed with the median of the training sample. 

After splitting the sample into training and validation sets (4:1 ratio), we train three independent regression models using different algorithms: XGBoost, TabNet \citep{Arik_Pfister_2021}, and FT-Transformer \citep{gorishniy2021revisiting}. Each model is trained with its optimal hyperparameters found by optuna. The final ensemble redshift $z_{\rm ph}$ is obtained by averaging the predictions of the three models. 

\begin{deluxetable}{ccccc}
\tablecaption{Scores of all photometric redshift regression models (XGBoost, TabNet, FT-Transformer, and the ensemble model) on the validation set. \label{tab:photoz_metrics}}
\tablehead{\colhead{Metric} & \colhead{XGBoost} & \colhead{TabNet} & \colhead{FT-Transformer} & \colhead{Ensemble}}
\startdata
RMSE & 0.2370 & 0.2329 & 0.2320 & 0.2256 \\
$\sigma_{\mathrm{NMAD}}$ & 0.0225 & 0.0212 & 0.0224 & 0.0187 \\
$f_c$ & 7.68\% & 7.13\% & 6.91\% & 6.92\%
\enddata
\end{deluxetable}

\begin{figure*}[ht]
  \centering
  \includegraphics[width=1\textwidth]{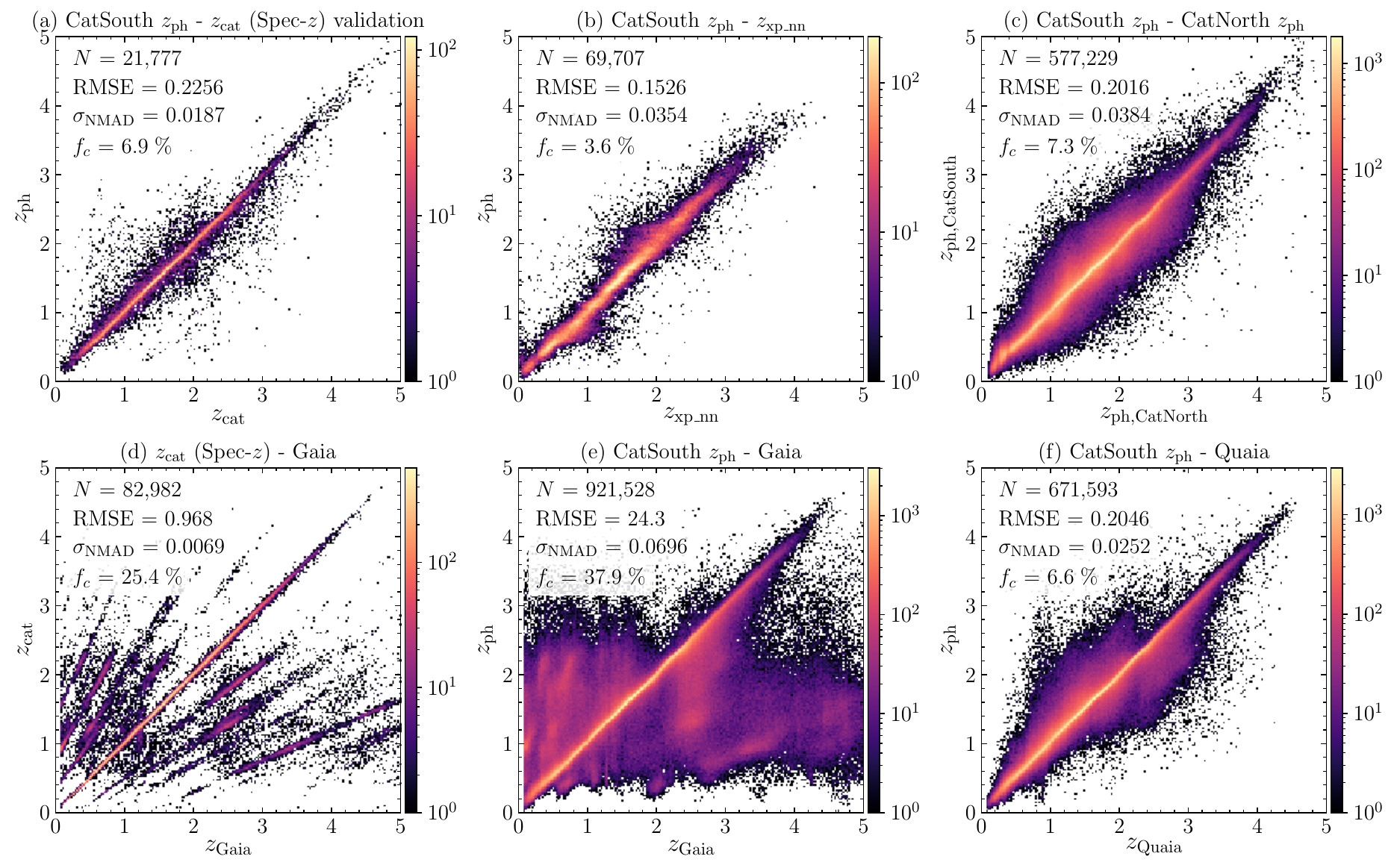}
\caption{Top row: ensemble photometric redshift ($z_\mathrm{ph}$) against spectral redshift ($z_\mathrm{cat}$) of the validation set with 21,777 quasars (a), $z_\mathrm{ph}$ against RegNet redshift ($z_{\mathrm{xp\_nn}}$) (b), and $z_\mathrm{ph}$ of CatSouth versus that in CatNorth for sources in common (c).
Bottom row: comparisons between the spectral redshift ($z_\mathrm{cat}$) in the training/validation sample and the \gaia\ redshift (d), CatSouth $z_\mathrm{ph}$ and \gaia\ (e), and CatSouth $z_\mathrm{ph}$ and Quaia (f). The plots are color-coded with two-dimensional densities (number counts in the pixels) of the samples, the values of which are indicated in the colorbars.}
\label{fig:redshift_validation_multi}
\end{figure*}

The scores of the three regression models and the ensemble model on a validation set of 21,777 sources are listed in Table \ref{tab:photoz_metrics}. Among the three base models, FT-Transformer achieves the lowest RMSE (0.2320) and $f_{\mathrm{c}}$ (6.91\%), and TabNet achieves the lowest $\sigma_{\mathrm{NMAD}}$ (0.0212). Averaging the three base models produces an ensemble model with even lower RMSE (0.2256) and $\sigma_{\mathrm{NMAD}}$ (0.0187), and a moderately low $f_{\mathrm{c}}$ (6.92\%). Because ensemble models {{not only reduce over-fitting (variance) but also lower the estimation bias \citep[see e.g.][]{Sagi2018}}}, we expect the ensemble model to be more robust than the individual base models.  Figure \ref{fig:redshift_validation_multi} shows the performance of the redshift regression models on the validation sets, and the comparisons between CatSouth redshift estimates and those from the GDR3 QSO candidate catalog, the CatNorth catalog, and the Quaia catalog.

% \section{Results} 
\section{Results: The CatSouth and CatGlobe Quasar Candidate Catalogs} \label{sec:results}

We compile the CatSouth quasar candidate catalog by including the \gdr{3}~photometry and astrometry, SMSS DR4 photometry, near-infrared photometry from VISTA surveys, WISE photometry from CatWISE2020 and AllWISE, and derived quantities (source probabilities, photometric and spectroscopic redshifts) from this work. Descriptions of the format of the catalog are shown in Table \ref{tb:catsouth_meta}. 

\begin{figure*}[ht]
    \centering
    \includegraphics[width=1\textwidth]{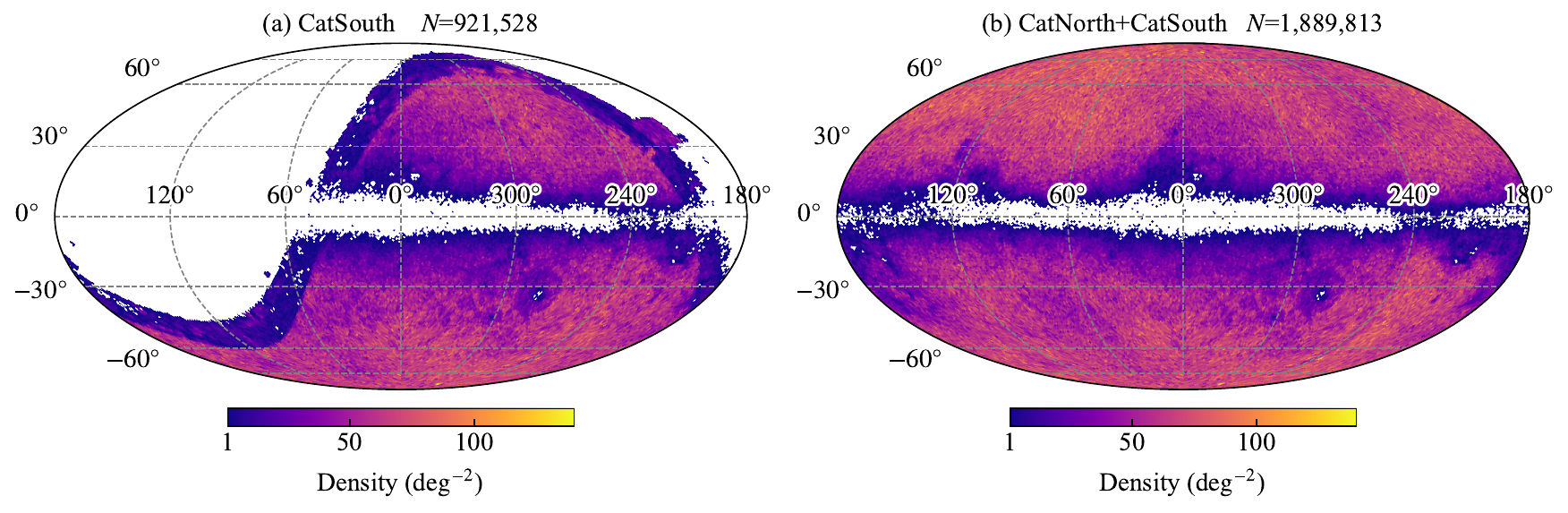}
    \caption{HEALPix \citep{2005ApJ...622..759G} sky density maps of the CatSouth quasar candidate catalog (a), and the CatGlobe (CatNorth+CatSouth) quasar candidate catalog (b). The maps are plotted in Galactic coordinates, with parameter $N_{\mathrm{side}}=64$ and an area of 0.839 $\mathrm{deg}^{2}$ per pixel.}
    \label{fig:sky_density}
\end{figure*}

The sky density distribution of CatSouth sources is shown in Figure \ref{fig:sky_density}. The low density around LMC and SMC is due to the additional source filtering described in Section \ref{sec:pm_filter}. Apart from the Galactic plane and the Magellanic Clouds, regions with $\delta \gtrsim 0\degr$ show low source density because of the lack of sky coverage of SMSS \citep[see Section 6.3 of][]{2024PASA...41...61O}. 

The median sky density of CatSouth is 41.7 $\deg^{-2}$, which is approximately $2/3$ of that of CatNorth (61.96 $\deg^{-2}$). 
This difference in sky density is mainly because the PS1 DR1 has a 1-to-2 magnitude deeper detection limit than SMSS DR4, and CatNorth is built on the former. CatNorth and CatSouth have 577,229 sources in common, which are mainly located at $-30 \degr \lesssim \delta\lesssim 16\degr$. Despite that CatNorth is more complete than CatSouth, CatSouth contains 14,543 objects in the PS1 footprint that are not in CatNorth. In addition, crossmatching CatSouth with the full Milliquas v8 using a radius of 1\farcs5 gives 128,596 sources in common, leaving 792,940 sources new to Milliquas. Such inclusion of new candidates indicates that different surveys and selection methods complement each other in building a more complete quasar sample.  

\startlongtable
\begin{deluxetable*}{p{0.05\textwidth}p{0.19\textwidth}p{0.06\textwidth}p{0.07\textwidth}p{0.53\textwidth}}
\centerwidetable
\tablewidth{1\textwidth}
\tabletypesize{\scriptsize}
\tablecaption{Format of the CatSouth quasar candidate catalog. \label{tb:catsouth_meta}}
\tablehead{{Column} & {Name} & {Type} & {Unit} & {Description}}
\startdata
1 & source\_id & long & ... & \gdr{3} unique source identifier \\
2 & ra & double & deg & \gdr{3} right ascension (ICRS) at Ep=2016.0 \\
3 & dec & double & deg & \gdr{3} declination (ICRS) at Ep=2016.0 \\
4 & l & double & deg & Galactic longitude \\
5 & b & double & deg & Galactic latitude \\
6 & parallax & double & mas & Parallax \\
7 & parallax\_error & double & mas & Standard error of parallax \\
8 & pmra & float & mas/yr & Proper motion in right ascension direction \\
9 & pmra\_error & float & mas/yr & Standard error of pmra \\
10 & pmdec & float & mas/yr & Proper motion in declination direction \\
11 & pmdec\_error & float & mas/yr & Standard error of pmdec \\
12 & pmra\_pmdec\_corr & float & ... & Correlation between pmra and pmdec \\
13 & phot\_bp\_mean\_mag & float & mag & Integrated BP mean magnitude \\
14 & phot\_g\_mean\_mag & float & mag & G-band mean magnitude \\
15 & phot\_rp\_mean\_mag & float & mag & Integrated RP mean magnitude \\
16 & bp\_rp & float & mag & BP$-$RP color \\
17 & phot\_bp\_rp\_excess\_factor & float & ... & BP/RP excess factor \\
18 & smss\_id & long & ... & SMSS DR4 unique object id \\
19 & ra\_smss & double & deg & SMSS DR4 R.A. (ICRS) \\
20 & dec\_smss & double & deg & SMSS DR4 decl. (ICRS) \\
21 & chi2\_psf & float & ... & Maximum chi-squared from photometry table \\
22 & g\_psf & float & mag & Weighted mean SMSS $g$-band PSF magnitude \\
23 & e\_g\_psf & float & mag & Error in g\_psf \\
24 & r\_psf & float & mag & Weighted mean SMSS $r$-band PSF magnitude \\
25 & e\_r\_psf & float & mag & Error in r\_psf \\
26 & i\_psf & float & mag & Weighted mean SMSS $i$-band PSF magnitude \\
27 & e\_i\_psf & float & mag & Error in i\_psf \\
28 & z\_psf & float & mag & Weighted mean SMSS $z$-band PSF magnitude \\
29 & e\_z\_psf & float & mag & Error in z\_psf \\
30 & yapermag3 & float & mag & Default point source $Y$ aperture corrected Vega mag (2\farcs0 diameter) \\
31 & yapermag3err & float & mag & Error in yapermag3 \\
32 & japermag3 & float & mag & Default point source $J$ aperture corrected Vega mag (2\farcs0 diameter) \\
33 & japermag3err & float & mag & Error in japermag3 \\
34 & hapermag3 & float & mag & Default point source $H$ aperture corrected Vega mag (2\farcs0 diameter) \\
35 & hapermag3err & float & mag & Error in hapermag3 \\
36 & ksapermag3 & float & mag & Default point source \ks aperture corrected Vega mag (2\farcs0 diameter) \\
37 & ksapermag3err & float & mag & Error in ksapermag3 \\
38 & catwise\_id & string & ... & CatWISE2020 source id \\
39 & ra\_cat & double & deg & CatWISE2020 R.A. (ICRS) \\
40 & dec\_cat & double & deg & CatWISE2020 decl. (ICRS) \\
41 & pmra\_cat & float & arcsec/yr & CatWISE2020 proper motion in right ascension direction \\
42 & pmdec\_cat & float & arcsec/yr & CatWISE2020 proper motion in declination direction \\
43 & e\_pmra\_cat & float & arcsec/yr & Uncertainty in pmra\_cat \\
44 & e\_pmdec\_cat & float & arcsec/yr & Uncertainty in pmdec\_cat \\
45 & snrw1pm & float & ... & Flux S/N ratio in band-1 (W1) \\
46 & snrw2pm & float & ... & Flux S/N ratio in band-2 (W2) \\
47 & snrw3 & float & ... & Flux S/N ratio in band-3 (W3) from AllWISE \\
48 & w1mpropm & float & mag & WPRO magnitude in band-1 (Vega) \\
49 & e\_w1mpropm & float & mag & Uncertainty in w1mpropm \\
50 & w2mpropm & float & mag & WPRO magnitude in band-2 (Vega) \\
51 & e\_w2mpropm & float & mag & Uncertainty in w2mpropm \\
52 & w3mpro & float & mag & WPRO magnitude in band-3 from AllWISE (Vega) \\
53 & e\_w3mpro & float & mag & Uncertainty in w3mpro from AllWISE \\
54 & phot\_bp\_rp\_excess\_factor\_c & float & ... & Corrected phot\_bp\_rp\_excess\_factor \\
55 & log\_fpm0 & float & ... & Logarithm probability density of zero proper motion ($\log f_\mathrm{{PM0}}$) \\
56 & in\_lmc & boolean & ... & Set to \texttt{True} if within 10\degr~from the center of LMC; \texttt{False} otherwise \\
57 & in\_smc & boolean & ... & Set to \texttt{True} if within 5\degr~from the center of SMC; \texttt{False} otherwise \\
58 & z\_gaia & float & ... & Redshift estimate from Gaia DR3 QSO candidate table \\
59 & z\_gaia\_low & float & ... & lower confidence interval of z\_gaia taken at 0.15866 quantile \\
60 & z\_gaia\_up & float & ... & Upper confidence interval of z\_gaia taken at 0.84134 quantile \\
61 & p\_gal & float & ... & XGBoost probability of the object being a galaxy \\
62 & p\_qso & float & ... & XGBoost probability of the object being a quasar \\
63 & p\_star & float & ... & XGBoost probability of the object being a star \\
64 & z\_ph\_xgb & float & ... & Photometric redshift predicted with XGBoost \\
65 & z\_ph\_tab & float & ... & Photometric redshift predicted with TabNet \\
66 & z\_ph\_ftt & float & ... & Photometric redshift predicted with FT-Transformer \\
67 & z\_ph & float & ... & Ensemble photometric redshift (mean of z\_ph\_xgb, z\_ph\_tab, and z\_ph\_ftt) \\
68 & z\_xp\_nn & float & ... & Spectral redshift predicted with RegNet using \gaia\ low-res spectroscopy
\enddata
\tablecomments{This table is published in its entirety in the machine-readable format. This table is also available on the PaperData Repository of the National Astronomical Data Center of China at doi:\href{https://doi.org/10.12149/101575}{10.12149/101575}.}
\end{deluxetable*}

\begin{figure}[ht]
    \centering
    \includegraphics[width=1\linewidth]{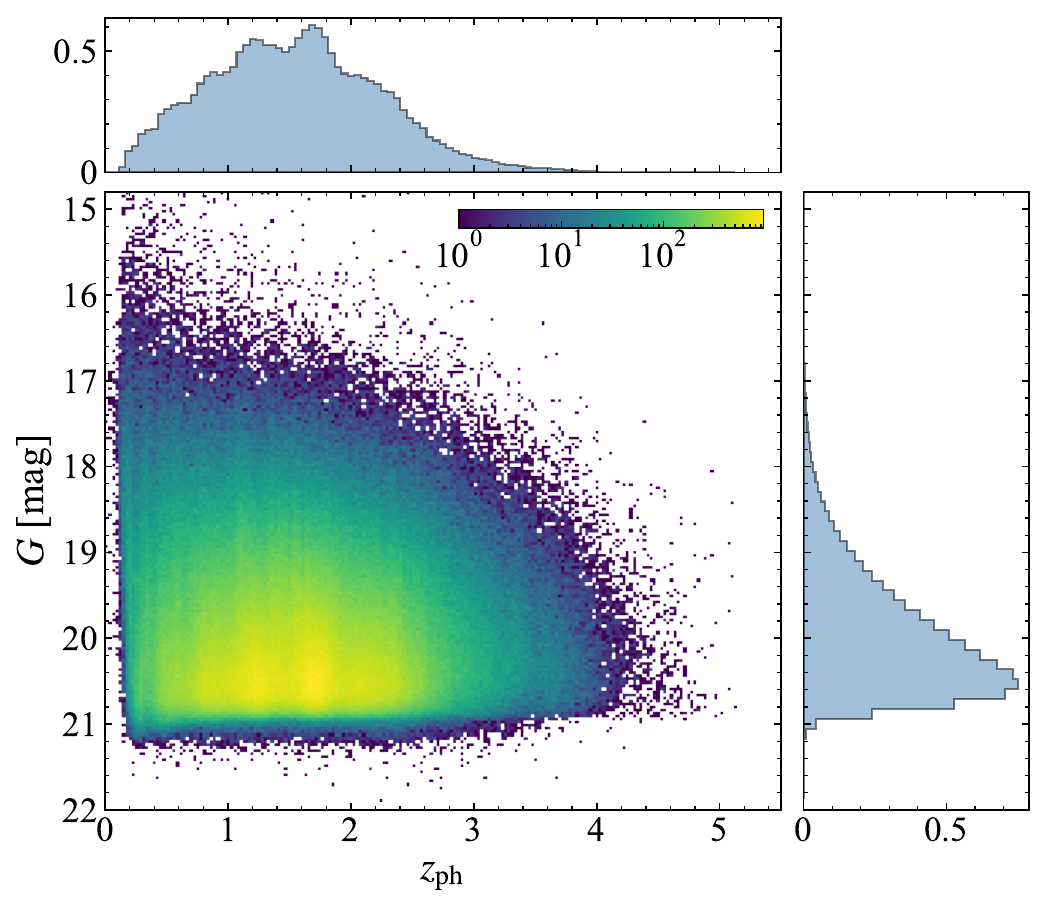}
    \caption{Apparent $G$ magnitude and photometric redshift distribution of CatGlobe quasar candidates. The $G$ magnitude is not extinction-corrected.}
    \label{fig:gmag_zph}
\end{figure}

By combining CatNorth and CatSouth, we generate ``CatGlobe'', a unified all-sky quasar candidate catalog based on \gdr{3}. We keep the CatNorth entry in the CatGlobe catalog when a source is in both CatNorth and CatSouth because CatNorth offers better depth. The final CatGlobe catalog includes 1,889,813 unique sources, whose table description is listed in Table \ref{tb:catglobe_meta}, and sky density distribution is shown in Figure \ref{fig:sky_density}. The distributions of the apparent $G$ magnitudes and photometric redshifts ($z_{\mathrm{ph}}$) of the CatGlobe quasar candidates are displayed in Figure \ref{fig:gmag_zph}. The photometric redshift ranges from 0 to approximately 5. Similar to the CatNorth catalog, the CatSouth and CatGlobe catalogs probe the bright end of quasars.

\begin{figure*}[htbp]
    \centering
    \includegraphics[width=1\textwidth]{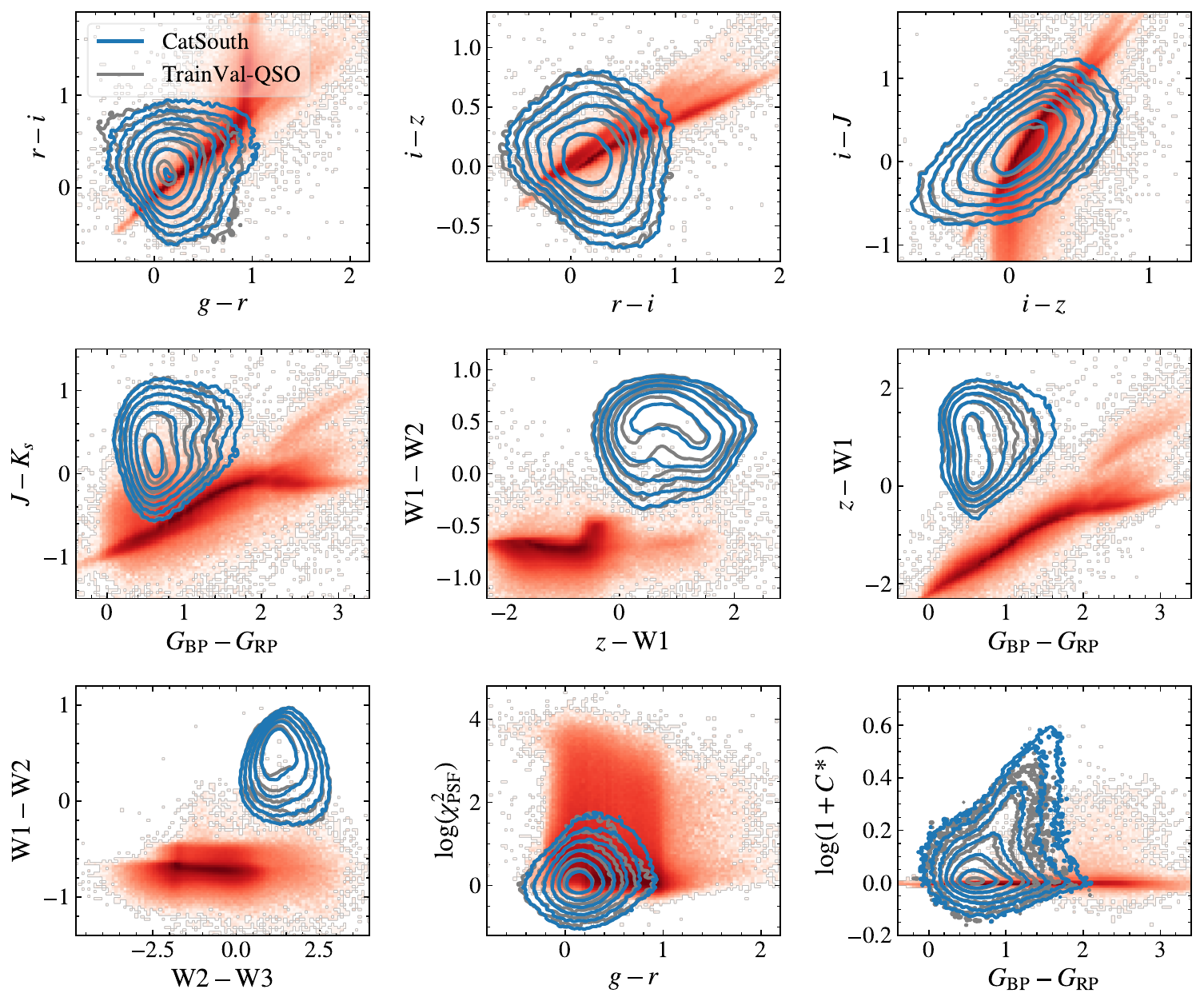}
    \caption{Two-dimensional feature representations (color-color and morphology-color diagrams) of sources in the CatSouth quasar candidate catalog (blue contours), quasars from the training/validation sample (gray contours), and stars from the training/validation sample (red-shaded density plots). To avoid clutter in the figures, galaxies are not plotted. All magnitudes are in the AB system and not dereddened.}
    \label{fig:ccd_catsouth}
\end{figure*}

As has been in Figure 14 of \citet{2024ApJS..271...54F} for CatNorth, and Figure \ref{fig:ccd_catsouth} in this work for CatSouth, our machine learning classification method recovers a quasar candidate sample with color-color and morphology-color distributions that are well-aligned with the bona fide quasar samples from the training/validation sets. 

\begin{figure*}[ht]
\centering
\begin{subfigure}{0.5\textwidth}
  \centering
  \includegraphics[width=\linewidth]{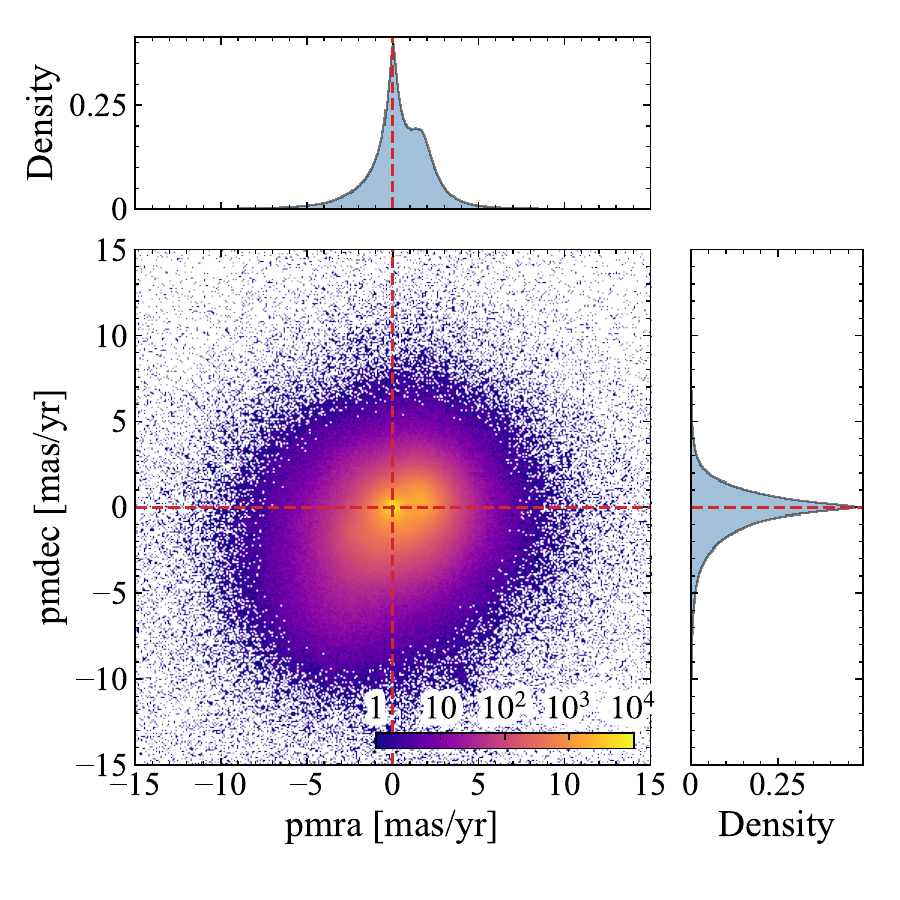}
\end{subfigure}%
\begin{subfigure}{0.5\textwidth}
  \centering
  \includegraphics[width=\linewidth]{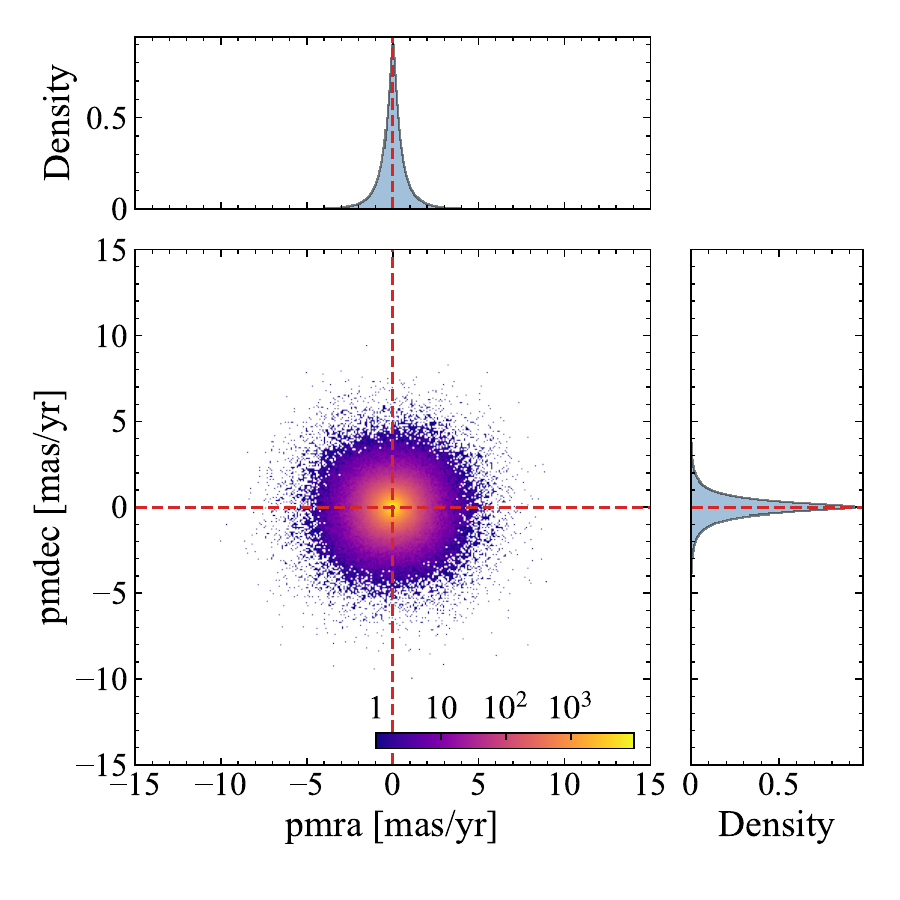}
\end{subfigure}
\caption{Left: joint and marginal density distributions of proper motions of the original GDR3 QSO candidates in right ascension and declination directions. The two-dimensional joint density plot is color-coded with the number of sources. Right: same as the left panel, but for CatGlobe quasar candidates from this work.}
\label{fig:pm2dhist}
\end{figure*}

In addition, we examine the proper motion distributions of the original GDR3 QSO candidate sample and the CatGlobe catalog in Figure \ref{fig:pm2dhist}. The original GDR3 QSO candidate sample shows asymmetric proper motion distributions in both right ascension and declination directions, many sources with relatively large proper motions (e.g., pmra or pmdec greater than 10 mas/yr), and overdensities in off-zero regions. Such behavior indicates modestly high stellar contamination in the GDR3 QSO candidates. In contrast, the CatGlobe quasar candidates show highly symmetric proper motion distributions in both directions, with only a small number of sources showing large proper motions. The improvement in the proper motion distributions suggests a significant enhancement of the purity of quasars in CatGlobe over the original GDR3 QSO candidates.

\begin{deluxetable*}{p{0.05\textwidth}p{0.19\textwidth}p{0.06\textwidth}p{0.07\textwidth}p{0.53\textwidth}}
\centerwidetable
\tablewidth{1\textwidth}
\tabletypesize{\scriptsize}
\tablecaption{Format of the CatGlobe quasar candidate catalog. \label{tb:catglobe_meta}}
\tablehead{{Column} & {Name} & {Type} & {Unit} & {Description}}
\startdata
1 & source\_id & long & ... & \gdr{3} unique source identifier \\
2 & ra & double & deg & \gdr{3} right ascension (ICRS) at Ep=2016.0 \\
3 & dec & double & deg & \gdr{3} declination (ICRS) at Ep=2016.0 \\
4 & l & double & deg & Galactic longitude \\
5 & b & double & deg & Galactic latitude \\
6 & parallax & double & mas & Parallax \\
7 & parallax\_error & double & mas & Standard error of parallax \\
8 & pmra & float & mas/yr & Proper motion in right ascension direction \\
9 & pmra\_error & float & mas/yr & Standard error of pmra \\
10 & pmdec & float & mas/yr & Proper motion in declination direction \\
11 & pmdec\_error & float & mas/yr & Standard error of pmdec \\
12 & pmra\_pmdec\_corr & float & ... & Correlation between pmra and pmdec \\
13 & phot\_bp\_mean\_mag & float & mag & Integrated BP mean magnitude \\
14 & phot\_g\_mean\_mag & float & mag & G-band mean magnitude \\
15 & phot\_rp\_mean\_mag & float & mag & Integrated RP mean magnitude \\
16 & bp\_rp & float & mag & BP$-$RP color \\
17 & phot\_bp\_rp\_excess\_factor & float & ... & BP/RP excess factor \\
18 & catwise\_id & string & ... & CatWISE2020 source id \\
19 & ra\_cat & double & deg & CatWISE2020 R.A. (ICRS) \\
20 & dec\_cat & double & deg & CatWISE2020 decl. (ICRS) \\
21 & snrw1pm & float & ... & Flux S/N ratio in band-1 (W1) \\
22 & snrw2pm & float & ... & Flux S/N ratio in band-2 (W2) \\
23 & w1mpropm & float & mag & WPRO magnitude in band-1 (Vega) \\
24 & e\_w1mpropm & float & mag & Uncertainty in w1mpropm \\
25 & w2mpropm & float & mag & WPRO magnitude in band-2 (Vega) \\
26 & e\_w2mpropm & float & mag & Uncertainty in w2mpropm \\
27 & phot\_bp\_rp\_excess\_factor\_c & float & ... & Corrected phot\_bp\_rp\_excess\_factor \\
28 & log\_fpm0 & float & ... & Logarithm probability density of zero proper motion ($\log f_\mathrm{{PM0}}$) \\
29 & z\_gaia & float & ... & Redshift estimate from Gaia DR3 QSO candidate table \\
30 & p\_gal & float & ... & XGBoost probability of the object being a galaxy \\
31 & p\_qso & float & ... & XGBoost probability of the object being a quasar \\
32 & p\_star & float & ... & XGBoost probability of the object being a star \\
33 & z\_ph\_xgb & float & ... & Photometric redshift predicted with XGBoost \\
34 & z\_ph\_tab & float & ... & Photometric redshift predicted with TabNet \\
35 & z\_ph\_ftt & float & ... & Photometric redshift predicted with FT-Transformer \\
36 & z\_ph & float & ... & Ensemble photometric redshift (mean of z\_ph\_xgb, z\_ph\_tab, and z\_ph\_ftt) \\
37 & z\_xp\_nn & float & ... & Spectral redshift predicted with RegNet using \gaia\ low-res spectroscopy
\enddata
\tablecomments{This table is published in its entirety in the machine-readable format. This table is also available on the PaperData Repository of the National Astronomical Data Center of China at doi:\href{https://doi.org/10.12149/101575}{10.12149/101575}.}
\end{deluxetable*}

\section{Summary and Conclusions}\label{sec:conc}

In this paper, we present the CatSouth quasar candidate catalog, an improved Gaia DR3 quasar candidate catalog in the southern sky. By combining Gaia DR3 astrometry and photometry with complementary data from SkyMapper DR4, VISTA surveys, and CatWISE2020, we implement a machine-learning classification selection method that effectively purifies the original Gaia quasar candidate catalog. We construct robust training/validation sets using spectroscopically confirmed quasars and high-quality CatNorth sources whose CNN-derived redshifts closely agree with the original Gaia estimates. With a set of carefully selected photometric and morphological features, the XGBoost classifier produces a high-purity sample of quasar candidates in the southern regions.

For quasar candidates with available Gaia BP/RP spectra, we directly derive spectroscopic redshifts using the pre-trained convolutional neural network (RegNet) from CatNorth. We train an ensemble photometric redshift model for the full sample based on XGBoost, TabNet, and FT-Transformer algorithms. Our ensemble photometric redshifts demonstrate competitive performance, with validation metrics indicating a significant improvement over the original Gaia redshifts and high consistency with the CNN-based spectroscopic redshifts.

The CatSouth catalog has limiting magnitudes of approximately $G\lesssim21$ and $i\lesssim21$. CatSouth has a median sky density of 41.7 $\deg^{-2}$, which is lower than that of CatNorth (61.96 $\deg^{-2}$) due to the shallower depth of SMSS DR4 than PS1. Nevertheless, CatSouth complements CatNorth and other existing quasar catalogs, especially in the southern hemisphere. By merging CatSouth with CatNorth, we produce the unified all-sky CatGlobe catalog, which provides a valuable resource for spectroscopic follow-up surveys, cosmological studies, and the construction of future Gaia celestial reference frames.

Our results demonstrate the effectiveness of combining multiwavelength data and advanced machine-learning techniques in the selection and redshift estimation of quasar candidates. The CatSouth catalog extends the quasar candidate sample to the southern hemisphere and improves the overall quality and reliability of redshift estimates, paving the way for future spectroscopic campaigns and enhanced celestial reference frame construction.

% Also note that the acknowledgment environment does not support long amounts of text. If you have a lot of people and institutions to acknowledge, do not use this command. Instead, create a new \section{Acknowledgments}.
% \begin{acknowledgments}
% \acknowledgments
% \section{Acknowledgments}
% \small
\begin{acknowledgments}
We acknowledge the support of the National Key R\&D Program of China (2022YFF0503401). We thank the support from the National Science Foundation of China (12133001) and the science research grant from the China Manned Space Project with No. CMS-CSST-2021-A06. The work is supported by the High-Performance Computing Platform of Peking University. We thank the referee for helpful suggestions to improve this paper.

KIC acknowledges funding from the Dutch Research Council (NWO) through the award of the Vici Grant VI.C.212.036 and funding from the Netherlands Research School for Astronomy (NOVA).
This work has made use of data from the European Space Agency (ESA) mission {\it Gaia} (\url{https://www.cosmos.esa.int/gaia}), processed by the {\it Gaia} Data Processing and Analysis Consortium (DPAC, \url{https://www.cosmos.esa.int/web/gaia/dpac/consortium}). Funding for the DPAC has been provided by national institutions, in particular the institutions participating in the {\it Gaia} Multilateral Agreement.
This publication uses data products from the SkyMapper Southern Survey data releases. The national facility capability for SkyMapper has been funded through ARC LIEF grant LE130100104 from the Australian Research Council, awarded to the University of Sydney, the Australian National University, Swinburne University of Technology, the University of Queensland, the University of Western Australia, the University of Melbourne, Curtin University of Technology, Monash University and the Australian Astronomical Observatory. SkyMapper is owned and operated by The Australian National University's Research School of Astronomy and Astrophysics. The survey data were processed and provided by the SkyMapper Team at ANU. The SkyMapper node of the All-Sky Virtual Observatory (ASVO) is hosted at the National Computational Infrastructure (NCI). Development and support of the SkyMapper node of the ASVO has been funded in part by Astronomy Australia Limited (AAL) and the Australian Government through the Commonwealth's Education Investment Fund (EIF) and National Collaborative Research Infrastructure Strategy (NCRIS), particularly the National eResearch Collaboration Tools and Resources (NeCTAR) and the Australian National Data Service Projects (ANDS).
This research uses services or data provided by the Astro Data Lab, which is part of the Community Science and Data Center (CSDC) Program of NSF NOIRLab. NOIRLab is operated by the Association of Universities for Research in Astronomy (AURA), Inc. under a cooperative agreement with the U.S. National Science Foundation.
This publication uses data products from the Wide-field Infrared Survey Explorer, a joint project of the University of California, Los Angeles, and the Jet Propulsion Laboratory/California Institute of Technology, funded by the National Aeronautics and Space Administration.
Funding for the Sloan Digital Sky Survey V has been provided by the Alfred P. Sloan Foundation, the Heising-Simons Foundation, the National Science Foundation, and the Participating Institutions. SDSS acknowledges support and resources from the Center for High-Performance Computing at the University of Utah. The SDSS web site is \url{www.sdss.org}. SDSS is managed by the Astrophysical Research Consortium for the Participating Institutions of the SDSS Collaboration, including the Carnegie Institution for Science, Chilean National Time Allocation Committee (CNTAC) ratified researchers, the Gotham Participation Group, Harvard University, Heidelberg University, The Johns Hopkins University, L’Ecole polytechnique f\'{e}d\'{e}rale de Lausanne (EPFL), Leibniz-Institut f\"{u}r Astrophysik Potsdam (AIP), Max-Planck-Institut f\"{u}r Astronomie (MPIA Heidelberg), Max-Planck-Institut f\"{u}r Extraterrestrische Physik (MPE), Nanjing University, National Astronomical Observatories of China (NAOC), New Mexico State University, The Ohio State University, Pennsylvania State University, Smithsonian Astrophysical Observatory, Space Telescope Science Institute (STScI), the Stellar Astrophysics Participation Group, Universidad Nacional Aut\'{o}noma de M\'{e}xico, University of Arizona, University of Colorado Boulder, University of Illinois at Urbana-Champaign, University of Toronto, University of Utah, University of Virginia, Yale University, and Yunnan University.
This research has made use of the VizieR catalog access tool, CDS, Strasbourg Astronomical Observatory, France (DOI : 10.26093/cds/vizier).
\end{acknowledgments}


\vspace{5mm}
\facilities{\gaia, Skymapper, VISTA, WISE}


\software{astropy \citep{2013A&A...558A..33A,2018AJ....156..123A,2022ApJ...935..167A},
          astroquery \citep{2019AJ....157...98G},
          % corner.py \citep{2016JOSS....1...24F},
          dustmaps \citep{2018JOSS....3..695M},
          FT-Transformer \citep{gorishniy2021revisiting},
          GaiaXPy \citep{daniela_ruz_mieres_2023_7566303},
          % GNU Parallel \citep{tange_ole_2018_1146014},
          healpy \citep{2019JOSS....4.1298Z},
          HEALPix \citep{2005ApJ...622..759G},
          KDEpy \citep{tommy_odland_2018_2392268},
          MOCPy \citep{2022ivoa.spec.0727F,matthieu_baumann_2024_14205461},
          optuna \citep{akiba2019optuna},
          pandas \citep{mckinney-proc-scipy-2010,2022zndo...7093122T},
          pytorch \citep{Ansel_PyTorch_2_Faster_2024},
          % PyFOSC \citep{yuming_fu_2020_3915021},
          scikit-learn \citep{pedregosa2011scikit},
          TabNet \citep{Arik_Pfister_2021},
          TOPCAT \citep{2005ASPC..347...29T},
          XGBoost \citep{chen2016xgboost}.
}



\appendix
\section{ADQL queries for selecting \gdr{3} stellar samples} \label{adql:gaia}
Here we provide the ADQL queries for selecting \gdr{3} stellar samples from the \gaia Science Archive (\url{http://gea.esac.esa.int/archive/}).

\subsection{The \gdr{3} OBA sample} \label{adql:gaia-oba}
\small
\begin{verbatim}
SELECT gs.source_id, gs.ra, gs.dec, l, b, 
parallax, parallax_error, parallax_over_error,
pm, pmra, pmra_error, pmdec, pmdec_error, 
pmra_pmdec_corr, phot_g_mean_mag, 
phot_bp_mean_mag, phot_rp_mean_mag, 
phot_bp_rp_excess_factor, 
astrometric_excess_noise, 
astrometric_excess_noise_sig, 
astrometric_params_solved, 
ruwe, ipd_frac_multi_peak, 
s.vtan_flag, gs.teff_gspphot, 
gs.distance_gspphot, 
ap.teff_esphs, ap.teff_esphs_uncertainty, 
ap.spectraltype_esphs, ap.flags_esphs
FROM gaiadr3.gaia_source AS gs
JOIN gaiadr3.gold_sample_oba_stars 
  AS s USING (source_id)
JOIN gaiadr3.astrophysical_parameters 
  AS ap USING (source_id)
WHERE gs.dec < 16
AND phot_g_mean_mag > 8
AND ruwe < 1.4 
AND astrometric_params_solved = 31
AND parallax_over_error > 10
AND ipd_frac_multi_peak < 6 
AND phot_bp_n_blended_transits < 10
AND ap.teff_esphs > 7000
AND gs.classprob_dsc_combmod_star > 0.9 
AND s.vtan_flag = 0
\end{verbatim}

\subsection{The \gdr{3} FGKM sample} \label{adql:gaia-fgkm}
\small
\begin{verbatim}
SELECT gs.source_id, gs.ra, gs.dec, l, b, 
parallax, parallax_error, parallax_over_error,
pm, pmra, pmra_error, pmdec, pmdec_error, 
pmra_pmdec_corr, phot_g_mean_mag, 
phot_bp_mean_mag, phot_rp_mean_mag, 
phot_bp_rp_excess_factor, 
astrometric_excess_noise, 
astrometric_excess_noise_sig, 
astrometric_params_solved, 
ruwe, ipd_frac_multi_peak,  
gs.teff_gspphot, aps.teff_gspphot_marcs, 
aps.teff_gspphot_phoenix
FROM gaiadr3.gaia_source AS gs 
JOIN gaiadr3.astrophysical_parameters 
  AS ap USING (source_id)
JOIN gaiadr3.astrophysical_parameters_supp 
  AS aps USING (source_id)
WHERE gs.dec < 16 
AND phot_g_mean_mag > 8
AND ruwe < 1.4 
AND random_index BETWEEN 0 AND 450000000
AND astrometric_params_solved = 31
AND parallax_over_error > 15
AND ipd_frac_multi_peak < 6 
AND phot_bp_n_blended_transits < 10 
AND gs.teff_gspphot > 2500 
AND gs.teff_gspphot < 7500 
AND gs.distance_gspphot < 
  1000/(parallax-4*parallax_error) 
AND gs.distance_gspphot > 
  1000/(parallax+4*parallax_error) 
AND (gs.libname_gspphot='MARCS' 
  OR gs.libname_gspphot='PHOENIX')
AND ap.logposterior_gspphot > -4000 
AND gs.classprob_dsc_combmod_star > 0.9 
AND gs.mh_gspphot > -0.8 
AND ABS(aps.teff_gspphot_marcs - 
  aps.teff_gspphot_phoenix + 65) < 150 
AND radius_gspphot < 100 
AND mg_gspphot < 12 
AND phot_bp_n_obs > 19 
AND phot_rp_n_obs > 19 
AND phot_g_n_obs > 150 
\end{verbatim}
% \section{Appendix information}


\bibliography{catsouth}{}
\bibliographystyle{aasjournal}


\end{CJK*}
\end{document}